\documentclass[acmlarge]{acmart}
\makeatletter
\newcommand{\myconfshort}{\acmConference@shortname}
\newcommand{\myconffull}{\acmConference@name}
\newcommand{\myconfdate}{\acmConference@date}
\newcommand{\myconfloc}{\acmConference@venue}
\AtBeginDocument{
  \fancypagestyle{firstpagestyle}{
    \fancyhead{}%
    \fancyfoot[C]{}%
  }
  \fancyhf{}
  \fancyhead[LO]{\@headfootfont\shorttitle}%
  \fancyhead[RE]{\@headfootfont\@shortauthors}%
  \fancyhead[LE]{\@headfootfont\footnotesize \myconfshort, \myconfdate, \myconfloc}%
  \fancyhead[RO]{\@headfootfont\footnotesize \myconfshort, \myconfdate, \myconfloc}%
  \fancyfoot[C]{}%
}
\makeatother
\acmBooktitle{\conffull\@ (\confshort), \confdate, \confloc}

\AtBeginDocument{%
  }
\usepackage[inkscapelatex=false]{svg}
\usepackage{subfig}
\usepackage{graphicx}
\usepackage{array}
\usepackage{placeins}
\usepackage{float}
\usepackage{booktabs}
\usepackage{longtable}
\usepackage{caption}
\usepackage{soul}

\setcopyright{acmlicensed}
\copyrightyear{2026}
\acmYear{2026}
\setcopyright{cc}
\setcctype{by}
\acmConference[FAccT '26]{The 2026 ACM Conference on Fairness, Accountability, and Transparency}{June 25--28, 2026}{Montreal, QC, Canada}
\acmBooktitle{The 2026 ACM Conference on Fairness, Accountability, and Transparency (FAccT '26), June 25--28, 2026, Montreal, QC, Canada}
\acmDOI{10.1145/3805689.3812413}
\acmISBN{979-8-4007-2596-8/2026/06}

\begin{document}

%If your title is lengthy, you must define a short version to be used
%in the page headers, to prevent overlapping text. The \verb|title| command has a ``short title'' parameter:
\title[Privacy Without Remedy]{Privacy Without Remedy: An Assessment of Data Broker Compliance with California Privacy Law}

\author{Anna-Maria Gueorguieva}
\authornote{Equal contribution.}
\affiliation{%
 \institution{University of Washington} %should this be the full Information School institution? 
 \city{Seattle}
 \state{Washington}
 \country{USA}}
\email{agueorg@uw.edu}

\author{Jennifer King}
\authornotemark[1]
\affiliation{%
  \institution{Stanford Institute for Human-Centered Artificial Intelligence}
  \city{Stanford}
  \state{California}
  \country{USA}}
  \email{kingjen@stanford.edu}

\author{Apoorva Panidapu}
\authornotemark[1]
\affiliation{%
  \institution{Stanford University}
  \city{Stanford}
  \state{California}
  \country{USA}}
  \email{panidapu@stanford.edu}

\author{Daniel E. Ho}
\affiliation{%
  \institution{Stanford University}
   \city{Stanford}
  \state{California}
  \country{USA}}
  \email{dho@law.stanford.edu}

\renewcommand{\shortauthors}{Gueorguieva, King, Panidapu, and Ho}

\begin{abstract}
California's consumer privacy law is widely deemed to be the most protective in the United States, one of the few to expressly regulate third party entities that buy and sell consumer data (data brokers). We offer the first empirical assessment of data broker compliance with the 2018 California Consumer Privacy Act (CCPA) and the 2023 Delete Act, which requires data brokers to register with the state and report consumer rights request metrics annually. First, we demonstrate that only 9\% of 522 registered data brokers were fully compliant with transparency requirements after the the Delete Act took effect, although we do identify slight improvements over time. Second, we descriptively characterize wide heterogeneity across data brokers in the volume of consumer rights requests received, with many reporting none.  We bring in external business data to explore correlates associated with this variation, a challenge given the general lack of opacity into broker business practices. Third, in an audit of a sample of 250 data brokers’ consumer request processes, we find that 43\% make it impossible for consumers to exercise all privacy rights and 64\% introduce at least one design feature that creates substantial friction into the consumer request process. Last, we show how these deficiencies stem from the decentralization of compliance decisions to brokers themselves, enforcement limitations, and regulatory ambiguity. We articulate reforms that could improve consumer privacy, transparency in broker practices, and compliance with these laws.
\end{abstract}

%CCS concepts


%% Keywords. 

\maketitle

\section{Introduction}
The integration of technology, social media, and artificial intelligence across our daily lives raises  concerns about data privacy, specifically how our personal data is collected, shared, and sold by businesses. While many consumers are generally aware that their data is collected by the businesses (first parties) they interact with, fewer are knowledgeable about the wide-reaching third party data collection ecosystem that exists alongside the websites and apps they visit regularly. The companies that collect and exchange consumer data in this ecosystem are data brokers: per California law, companies that “knowingly” buy and sell the data of consumers with whom they do not have first-party relationships (Cal. Civ. Code § 1798.99.80). Data brokers can have over 10,000 different data types on offer for purchase collected from and inferred about consumers available for other businesses, individuals, and even governments \cite{altman-2024}. The industry is projected to be valued at \$462B by 2031 \cite{altman-2024}. In addition to being integral to not only the internet-wide consumer targeted advertising ecosystem, industries such as insurance, financial lenders, non-profits, real estate, and investigators, to name a few, purchase data from brokers for uses such as fraud detection and identify verification \cite{ftc-2014}. But data purchased from data brokers can also be used in harmful ways, such as for political violence \cite{ng-2025, us-court-2022}, national security threats such as identifying military personnel \cite{sherman-2023}, and targeted cases of identity theft \cite{casal-2024}.

The California Consumer Privacy Act (CCPA) of 2018 was the first U.S. state law to address data privacy protections for consumers. Described as the most “ambitious and comprehensive piece of privacy legislation” enacted at the state level to date \cite{blanke-2020}, the CCPA establishes rules for businesses' collection and use of consumer data, with the goals of increasing accountability and public transparency about these practices, as well as providing consumers with a set of data rights. Additionally, companies are required to post publicly on their privacy policy the number of rights requests that they received and fulfilled in the previous calendar year (Cal. Civ. Code § 1798.99.85). California's Delete Act of 2023 explicitly extends the transparency requirements to data brokers and also requires them to pay a fee and register with the California Privacy Protection Agency's (CPPA) Data Broker Registry, a \href{https://cppa.ca.gov/data_broker_registry/}{publicly accessible database}. 

Consumer advocacy groups rank California’s privacy law framework as the strongest in the country due to its breadth of consumer rights, transparency requirements, and enforcement structure \cite{epic-pirg}, calling it a ``historic, pro-consumer bill'' \cite{bracy-2023-iapp}. Commentators have also highlighted its effect beyond California in reshaping corporate data governance and establishing privacy compliance norms \cite{bloomberg-privacy, chander-2021}. While researchers have focused on evaluating both businesses’ compliance with and the effectiveness of the CCPA, research on data broker compliance with the CCPA or the Delete Act is limited. We hence investigate data broker compliance with two primary requirements in these Acts: public reporting requirements annually tracking consumers’ data rights requests metrics (i.e., the transparency requirement); and, an obligation to respond to consumer rights requests in compliance with the CCPA and without the use of misleading and obfuscating design (i.e., dark patterns) that subvert consumer will and creates friction with exercising these rights. Our findings provide relevant evidence to California policymakers and regulators \cite{calprivacy-2026} and provide greater opacity about data brokers' practices. %Currently, only four states have laws targeting data brokers. As more states develop regulations for data brokers and turn to California as a model, it is imperative that we first understand whether data brokers are complying with them.  

We make four research contributions. First, through manual collection and review of privacy policies from all 522 registered data brokers, we show that only 9\% of data brokers fully comply with reporting all six transparency requirements. Compliance increases with the type of request, e.g., 53\% of data brokers report the number of requests received for the two most widely asserted rights, the “do not sell” (opt-out) and the deletion rights. Second, inspection of the privacy request submission process shows that 43\% of brokers do not provide consumers with the ability to exercise all required data rights. In addition, 72\% of brokers engage in at least one behavior that is a violation of the CCPA, and 64\% of brokers have interface features that increase friction in the submission process. Third, we empirically characterize the correlates of whether data brokers comply, the number of requests received, and friction in fulfilling consumer requests. Fourth, we analyze how the decentralization of compliance decisions to data brokers has led to fragmented and inconsistent compliance and derive affirmative policy recommendations to improve the regulation of data brokers. 

%\section{Background and Related Work}
%In this section we start by reviewing the reporting requirements of the CCPA and the Delete Act. We then provide background information about data brokers and why California law seeks to regulate their practices. We conclude with a review of relevant literature that similarly seeks to evaluate data broker compliance to CCPA.

\section{Institutional and Legal Background on California Law}
\textbf{California Consumer Privacy Act.} Passed in 2018 and later updated by ballot initiative in 2020 (as the California Privacy Rights Act, or Proposition 24), the CCPA introduced requirements for businesses that collect, use, retain, and share consumer data. The law introduced a set of six data rights for California consumers: a right to delete one's data (Cal. Civ. Code § 1798.105); a right to correct personal information (Cal. Civ. Code § 1798.106); a right to know what personal information is being collected by a business (Cal. Civ. Code § 1798.110); a right to know what personal information is being sold/shared by a business (Cal. Civ. Code § 1798.115); a right to request the business not sell or share your data (a.k.a. “Do not sell”) (Cal. Civ. Code § 1798.120); and, a right to limit the use and disclosure of sensitive personal information (Cal. Civ. Code § 1798.121). %As of September 2025, updated CCPA regulations now require brokers to visibly indicate when an opt-out preference signal has been honored, expand the right to know by requiring businesses that retain personal information for longer than twelve months to provide consumers access to data collected as far back as January 1, 2022, and strengthen the right to limit by mandating that the Notice of the Right to Limit be delivered through the same mechanisms used to collect sensitive data. 
As the nation’s first state-level consumer data privacy law, California is catalyzing U.S. privacy law and has emerged as a ``privacy super-regulator'' \cite{chander-2021}. Since 2019, twenty-two additional states have passed consumer data privacy rights bills (of varying levels of strength) \cite{privacy-law-chart} and the CCPA has been the ``impetus behind those bills'' \cite{chander-2021}, increasing the pressure on Congress to act nationally to harmonize the various approaches to regulating data privacy. 

The CCPA requires businesses that collect the data of at least 10M California consumers to publicly post the number of rights requests they have fulfilled and the mean or median time to fulfill requests over the previous calendar year by July 1st, either within their website’s privacy policy or on a webpage linked directly from it (Cal. Civ. Code § 1798.99.85). The Attorney General’s (AG) statement of reasons for this requirement included that public rights reporting would ensure compliance with the law, help determine whether businesses were systematically denying consumer requests, and provide transparency specifically to “enable academics, consumer advocates, business groups, and others to research and analyze this data” \cite{ccpa-fsor}. The AG further noted that these transparency measures would aid in government enforcement and assist the public in exercising their rights. Reports of the number of rights requests received can include non-Californians (subsection (g)(4)), though businesses that choose to do so must disclose that their totals include non-residents, though they are not required to segment them from California-based requests. %However, many data brokers do not specify geographic boundaries of the requests reporting. The rationale for this decision was to make it easier for businesses to comply with reporting requirements, especially those that were already complying with similar rights requests from European data subjects under the General Data Protection Regulation (GDPR). As we will discuss later, this ambiguity adds complexity to our analysis, particularly with respect to those businesses that report millions of rights requests, making it unclear from whom these requests were received. 

\textbf{2023 Delete Act.} While the CCPA applies to data brokers as well as first-party data collectors, the Delete Act specifically broadens brokers' obligations. A public-facing data broker registry, originally administered by the California AG, was introduced in 2018 with the CCPA. The Delete Act moved the administration of the data broker registry to the CPPA (Cal. Civ. Code § 1798.99.82) and expanded rights reporting obligations for brokers by requiring them to comply with the CCPA’s rights request public posting requirement whether or not they met the 10M California consumer threshold for first party collectors (Cal. Civ. Code § 1798.99.85 (a)).\footnote{The Delete Act does not require brokers to report requests to correct information, a potential oversight, though they must honor requests to do so.} Importantly, registration reflects a broker’s past activity, not its future intent; brokers that began operating in 2023, for example, would not appear in the registry until filing for the first time in 2024. Additionally, data brokers must also indicate when they register whether they collect precise geolocation data, reproductive health data, and data from minors, and whether they are regulated by the Fair Credit Reporting Act (FCRA), the Gramm-Leach-Bliley Act (GBLA), the Insurance Information and Privacy Protection Act (IIPAA), or the Health Insurance Portability and Accountability Act (HIPAA). Governance by any of these federal statutes preempts some CCPA obligations.\footnote{As of 2026, SB 361 expanded these obligations to include: account login credentials, government issued ID numbers, citizenship and immigration status, union membership, sexual orientation, gender identity, and biometric data, among others. Brokers must also state whether they sell or share data to foreign actors, state and federal governments, law enforcement, and generative AI developers.}

The Delete Act also authorized the creation of a new technical platform, the Delete Request and Opt-out Platform (DROP), that allows California residents to submit data deletion and opt-out of sale requests to all registered data brokers with a single request \cite{ca-drop}. This Act expands the CCPA deletion right, which only applies to data collected from a consumer, while under DROP data brokers will have to delete all non-exempt personal information related to the consumer beyond basic identifiers, including behavioral, financial, health, location, and relationship data, as well as any inferences drawn about individuals from this data. Brokers will be required to begin honoring deletion requests as of August 2026 and must process requests every 45 days. Details regarding how the Agency will enforce compliance by the 500+ registered brokers are not public, though the Delete Act requires brokers to undergo third party audits assessing their compliance with the Act starting in 2028 and repeating every three years. Results must be provided to the Agency upon request. Remedy for a consumer who believes their rights under these Acts to be violated by a data broker (outside of a data breach) must report such to the CPPA,\footnote{https://cppa.ca.gov/webapplications/complaint} as there is no general private right of action.\footnote{CCPA has generally been interpreted to contain only a limited private right of action in the case of data breaches. But see Shah v.\ Capital One Financial Corp., 768 F.\ Supp.\ 3d 1033 (N.D.\ Cal.\ 2025) (finding that disclosure of information to third parties without consent may be sufficient to state a CCPA cause of action).}

\textbf{Data Brokers.} A data broker is defined as “a business that knowingly collects and sells to third parties the personal information of a consumer with whom the business does not have a direct relationship” (Cal. Civ. Code § 1798.99.80). By aggregating data collected from public records, mobile location, cross-site web, and mobile app tracking, many data brokers create detailed individual consumer profiles and generate ``consumer scores''—based on both actual and inferred data—that commodify personal information in ways opaque to the consumer \cite{crain-2017}. While broker-aggregated data is primarily used for consumer marketing purposes, such as identifying individuals for targeted advertisements \cite{us-senate-committee-2013}, it is increasingly used for other purposes, such as predictive analytics \cite{pasquale-2014}. In 2024, for example, General Motors was discovered to be selling their customers’ driving data without consent to two data brokers, who in turn used the data to create risk scoring products for auto insurance companies \cite{day-2025}. Numerous GM customers were denied auto insurance or found that their premiums increased on the basis of such scores. The company settled with the Federal Trade Commission and cannot sell individual customer data for five years.

Relying upon brokered data to classify or make predictions about individual consumers can lead to inaccurate, and sometimes discriminatory, decisions \cite{venkatadri-2019}. When background checks are used to evaluate individuals for jobs and housing opportunities, this data profiling can have disproportionately negative effects on minority groups, immigrants, and low-income residents \cite{ebadolahi-2023}. Individual data profiling can pose threats to the online and physical safety of consumers. For a small price, anyone can purchase detailed and personal information about an individual. For example, a week’s worth of data of visits to and from a Planned Parenthood clinic was sold for \$160 \cite{cox-2022}, allowing the purchaser to isolate the mobile device IDs of any visitor. A group of Colorado Catholics purchased mobile app location data to track clergy suspected of being gay, resulting in the outing of a priest in 2021 after his mobile location data showed he visited gay establishments \cite{boorstein-2021, boorstein-2023}. Multiple political and legal authorities in the U.S. have been assassinated or experienced attempted assassination by perpetrators who tracked them using profiles obtained from people search services \cite{adler-2025}. Given the potential for such harmful uses, evaluating the effectiveness of data broker-focused regulation is imperative. 

\section{Related Works}

Prior work investigating data broker compliance with the CCPA is sparse. Extant studies focus on a subset of brokers or one specific privacy right. In 2020, Consumer Reports examined ``do not sell'' provisions and found that ``data brokers’ processes for requesting opting out of selling your data are so onerous that they have substantially impaired consumers’ ability to opt out'' \cite{mahoney-2020}. Take et al. examined people search websites and found difficulties and complications with exercising the deletion right \cite{take-2024}. %In response to these challenges, third party agents authorized by the CCPA fill this gap by offering both free and paid services that process deletion and opt-out requests on behalf of individuals, though their services generally focus on only two rights. 
Researchers have documented that brokers often do not comply with consumers’ requests to know what data is being collected about them \cite{vankempen-2025}. Recent analysis on the effectiveness of data broker registries has found that across four state registries (including California's), hundreds of data brokers registered in one state are not registered in others \cite{trujillo-2025}. %This calls for additional insight into whether these discrepancies arise from noncompliance, definitional gaps, geographical scope, or other factors. 
To date, no studies have examined data broker compliance with the CCPA’s full set of obligations. 

Beyond California, the regulation of data brokers remains fragmented. Legal theorists have argued that data brokers' business models are incompatible with the General Data Protection Regulation (GDPR) in the European Union, which does not explicitly regulate data brokers as a category \cite{ruschemeier-2023}. Consequently, enforcement in the EU has largely occurred through case-by-case investigations by Data Protection Authorities with no systematic oversight as regulators continue to be stymied by the opacity of brokers' business practices \cite{hunton-2018, ren-2019}. 
Some academics have suggested that California's data broker registry could serve as a model for Europe to create increased visibility into the data broker marketplace \cite{ruschemeier-2024-shadows}. The increasing adoption of consumer privacy laws and data broker registries across US states---with four registries enacted and with several states developing similar frameworks---means that California provides a useful setting for empirically studying the effectiveness of transparency requirements in increasing accountability in the data broker ecosystem.

Other studies evaluate non-data broker companies compliance with the CCPA. These have largely focused on how opt-out links \cite{tran-2024, tran-2025, siebel-2022, oconnor-2021, charatan-2023}, interface design choices \cite{habib-2021} and the clarity of privacy policies \cite{chen-2021, hosseini-2024} can all contribute to inhibiting consumer privacy rights. We build on these insights to assess the privacy request processes of data brokers, which increase friction for consumers and may violate the CCPA.

\section{Research Design and Methods}

\textbf{Manual Assessment of Disclosures and Request Processes.} First, we measured compliance with the statutory transparency requirements by visiting and reviewing the websites of all 522 registered data brokers, verifying that each broker had a posted privacy policy, and then whether they reported rights requests. Brokers must post their rights requests either directly within their privacy policy or on a webpage linked directly from the privacy policy for the previous calendar year (e.g., 2025 postings must use 2024 calendar year data). We conducted this manual assessment to measure broker compliance with transparency requirements both before and after the July 1st, 2025 reporting deadline. 

Second, to measure compliance with the CCPA's requirements that submitting a consumer rights request must be a clear, easy-to-understand, and easy-to-execute process that does not utilize dark patterns (Cal. Code Regs. Tit. 11, § 7004(a)), we conducted an interface analysis using a stratified random sample of 250 of the 522 data brokers’ online request processes. We coded the brokers based on: (1) identification of noncompliant features of the interface, such as having a broken link or email; and, (2) specific features that increase friction in the submission process for consumers, such as having to submit multiple forms with the same personal information. 

%In our review of reporting metrics, we observed that there was substantial disparity between brokers in terms of the number of requests reported. We hypothesize that features of the business, such as income, type of data collected, and if it is a parent or subsidiary company, can impact the amount of requests received. We additionally hypothesize that companies with less features of noncompliance and friction in the request process have more requests. 

\textbf{Identifying Rights Requests Metrics.} The Delete Act requires that five metrics be publicly posted on a data broker’s privacy policy (\S~1798.99.82(b)(2)(B)): the total number of requests to (a) delete personal information; (b) know or access what personal information is collected; (c) know what personal information the data broker was selling or sharing and to whom; (d) opt out of sale or sharing of personal information; and (e) limit the data broker's use and disclosure of sensitive personal information. In addition to the request numbers, they must also report the mean and median number of days to fulfill each request type, and the number of requests they complied with and denied (in addition to number received). Finally, they must include their website URL, a URL to make data rights requests, mailing address, and any additional information they wish to provide. 

%The registry contains the consumer rights requests totals that brokers received in the prior calendar year (two years prior to the registry year); for example, the 2025 registry contains the 2023 metrics for consumer rights requests. Brokers are not required to publicly post the previous calendar year metrics until the deadline of July 1st, and then they report the same figures when submitting their registration in the following January. \hl{A timeline of these reporting deadlines is provided in Table}~\ref{tab:reporting_timeline} \hl{in Appendix} \ref{reporting-timeline}. 
As brokers are required to submit request reporting metrics when they register, we compared publicly reported metrics from privacy policies to the metrics brokers reported directly to the registry. However, metrics reported to the 2025 registry are from \emph{two} years prior (2023), rather than from 2024.\footnote{Table \ref{tab:reporting_timeline} in Appendix A presents further details on the timeline of requirements.} Additionally, missing values during registration were automatically replaced by the CPPA with zeros in registry data, making it difficult to determine which brokers did not report and which brokers actually received zero requests. Therefore, we could not verify exact matches between registry submissions and public postings, and instead use our manual assessment of privacy policies as our main measurement of compliance to transparency requirements.

%Because of the disjoint between registry metrics (2023 data) and public posts (2024 data), we did not rely on the registry metrics and instead evaluated transparency based on what brokers publicly posted on their privacy policies as required by the Delete Act.
To ascertain whether brokers would comply with the July 1st public reporting deadline, we visited and saved HTML copies of all broker privacy policies in June 2025 and then 45 days after the July 1st, 2025 deadline. The metrics reported on the privacy policies serve as our main source of data for ascertaining broker compliance with the transparency reporting requirement and for conducting analysis on the amount of requests data brokers receive.

\textbf{Measuring Compliance with Transparency Requirements.}
We developed a codebook based on CCPA and Delete Act regulations to analyze the privacy policies for compliance (see Appendix \ref{sec:codebook}). If a broker reported a specific request type, we documented the number of requests received, how many the broker complied with, and how many the broker denied. %Under the Delete Act, data brokers must report on their privacy policy webpage the number of requests received for: the right to know what data is being sold, what data is being collected, to delete your data, to opt-out of sales or sharing of data, and to limit sharing of sensitive personal information.
If a broker fails to report any one of the five metrics they are noncompliant with the Delete Act. We recorded “none” for a given right request amount if the data broker did not report it to differentiate from brokers who reported zero requests. Some data brokers may report the two rights to know in one metric,\footnote{Combining the two requests to know does not align with the CCPA as Sections 1798.110 and 1798.115 separate the right to know data being collected and sold, respectively. However, even the CPPA sometimes collapses these two rights into one in their own \href{https://cppa.ca.gov/meetings/materials/20250926_item4.pdf}{reports} (slide 5) \cite{cppa-enforcement-update-2025}.} so we recorded whether the data broker separated or grouped the two rights. If a data broker did not report the right to limit sensitive information but explicitly stated that they do not share sensitive information, we count the request to limit as being reported and document zero requests to limit for that broker.

We evaluated brokers’ compliance with the requirement to report specific rights requests both before the July 1st, 2025 Delete Act deadline and 45 days after. We then calculated reporting across all rights requests to determine brokers’ overall compliance with the reporting requirement. Summary statistics of the requests metrics are described in Table \ref{tab:descriptive-stats}; these preliminary findings motivated our further investigation into the request process and understanding the significant variation in request numbers.

\begin{table}[t]
    \centering
    \caption{Number of requests received by registered brokers as of August 2025. Source: data broker privacy policies.}
    \label{tab:descriptive-stats}
    \begin{tabular}{lrrrrr}
        \toprule
        & Mean & SD & Median & Max & \% brokers reporting \\
        \midrule
        Total Requests& 461{,}801 & 2{,}708{,}800 & 3{,}268 & 29{,}627{,}364 & 54\% \\
        Deletion Right
        & 39{,}740 & 260{,}548 & 7{,}037 & 4{,}063{,}777 & 53\% \\
        Do Not Sell Right
        & 429{,}132 & 2{,}723{,}323 & 30{,}265 & 29{,}615{,}957 & 53\% \\
        Right to Know (collecting)
        & 377 & 1{,}750 & 10 & 21{,}589 & 52\% \\
        Right to Know (selling)
        & 24 & 77 & 2 & 405 & 36\% \\
        Right to Limit
        & 1{,}452 & 5{,}737 & 0 & 29{,}673 & 36\% \\
        \bottomrule
    \end{tabular}
\end{table}

\begin{table}[t]
\centering
\caption{Percentages of data brokers collecting specific categories of sensitive data and regulation by federal laws. Source: California Data Broker Registry, 2025.}
\label{tab:self-reported-data}
\resizebox{\columnwidth}{!}{\begin{tabular}{lcccccccc}
\toprule
 & \shortstack[l]{Collects data\\from minors}
 & \shortstack[l]{Collects precise\\geolocation data}
 & \shortstack[l]{Collects reproductive\\health data}
 & \shortstack[l]{Subject to\\FCRA}
 & \shortstack[l]{Subject to\\GLBA}
 & \shortstack[l]{Subject to\\IIPPA}
 & \shortstack[l]{Subject to\\CMIA}
 & \shortstack[l]{Subject to\\HIPAA} \\
\midrule
\textbf{No}
 & 96\% & 84\% & 98\% & 96\% & 95\% & $>$99\% & 99\% & 94\% \\
\textbf{Yes}
 & 4\% & 16\% & 2\% & 4\% & 5\% & $<$1\% & 1\% & 6\% \\
\bottomrule
\end{tabular}}
\end{table}

\textbf{Evaluating Brokers’ Processes for Enabling Rights Requests.} The CCPA mandates that a “business shall not add unnecessary burden or friction to the process” (Cal. Code Regs. § 7004(a)(5)). In addition to intentional barriers introduced by design friction, policymakers were also concerned about the general use of dark patterns in rights requests. Dark patterns are defined in the statute as “user interfaces designed or manipulated with the substantial effect of subverting or impairing user autonomy, decisionmaking, or choice” (Cal. Civ. Code § 1798.140). The Delete Act requires that data brokers’ websites must provide a link to a page that enables consumers to exercise their privacy rights and that it should not “make use of any dark patterns” (Cal. Civ. Code § 1798.99.82(b)(2)(G)). 

We investigated whether data brokers complied with these design requirements by comparing the rights request interfaces against CCPA statutory requirements. Previous research identifying dark patterns in businesses’ compliance with the CCPA utilized qualitative coding to identify design features that increase friction in the consumer request process \cite{tran-2025}. We built upon that research by developing a taxonomy to classify design features of the rights requests interfaces, assessing whether they increase the difficulty, or friction, to exercise CCPA privacy rights. As an example of friction, Tran et al. \cite{tran-2025} identified the unnecessary insertion of CAPTCHA puzzles into the request process. If the link to a broker’s rights request process was broken we categorized it as not compliant (Cal. Code Regs. Tit. 11, § 7004). The design features we identified as noncompliant and increasing friction are included in the results (Table \ref{tab:rq2-results}). 

We reviewed the interfaces of 250 data brokers using stratified sampling based on the number of requests received in order to create a representative sample across request amounts, which were highly heterogeneous. We stratified into five mutually exclusive groups: data brokers that do not report metrics at all, and then four quartiles of reported requests. We randomly selected 50 brokers from those that did not report any request metrics (N=238) across all five rights request categories, which represents nearly half (46\%) of the total registered brokers. In order to sample from the remaining 284 brokers that did report metrics from at least one category, we sorted them by total requests received in 2024 and sampled 50 brokers in each quartile, where each quartile contained 71 data brokers in total.

\textbf{Understanding Variations in Requests.} As we document in Table \ref{tab:descriptive-stats}, there is wide heterogeneity across brokers in the number of requests they report. We investigate correlates of request volumes using: (a) self-reported attributes in the 2025 registry; and (b) merging data from Dun \& Bradstreet (D\&B) on corporate entity attributes (e.g., annual income, number of employees). We similarly explore correlates of noncompliance and friction in consumer request processes. Self-reported registry data capture whether a broker collects specific sensitive information (e.g., information from minors, reproductive health, and precise geolocation), if they are located in California, and whether the nature of their business obligates them to comply with other laws, such as the Fair Credit Reporting Act (FCRA); the percentage of brokers that collect sensitive information and are subject to these laws is listed in Table \ref{tab:self-reported-data}. %Using the self-reported data, we compare if the proportion of each of these variables that is true is significantly different between brokers that report vs don’t report, as well as lower requests receiving vs higher request receiving brokers. These features are then also used in the models described below. 
%
%\subsubsection{D\&B Establishment Level Data}
We use D\&B's Establishment Level Data database to infer income, number of employees, and corporate subsidiary status. The latter is relevant, as we observed similarities in privacy policies for related parties. We successfully matched 239 data brokers by company name (or secondary name) and ZIP code. A full description of the predictor variables is available in the Appendix \ref{sec:predictor-variables}. We conducted this analysis by comparing features across low and high request reporting data brokers, splitting into two groups by the median request value. We additionally use logistic regression and analysis of variance to investigate the features associated with whether or not data brokers report metrics at all.

%We gathered data on subsidiary structure as review of privacy policies showed implicit similarities across brokers without formal relationships, leading us to assume that some brokers may be subsidiaries of other brokers or companies that could impact the amount of requests they receive. Ultimately, 
 %We run an exploratory analysis of the difference in means between these variables for reporting vs not reporting brokers, as well as for lower requests receiving vs higher request receiving brokers. We combined this data with the self-reported registry data to conduct exploratory modeling of the relationships described below. 
%\subsubsection{Predictive Modeling}
 %using the same response variables to predict five different response variables: the total requests received, and for those brokers that report this data, the number of requests to opt out received, requests to delete received, requests to know received, and requests to limit received. %Results are described in Section 4.3. 

\section{Results}
%Our analysis reveals that data brokers are weakly compliant with rights request reporting. Noncompliance and friction are prevalent across brokers’ request submission processes. We also identified that the type of data they collect, their annual income, and the amount of friction we found in their rights requests processes are associated with a lack of compliance with reporting requests and predictive of which types of brokers receive greater numbers of requests from consumers. 

\begin{figure}[!t]
  \centering
  \begin{minipage}[t]{0.48\columnwidth}
  \vspace{0pt}
    \centering
      \includegraphics[width=\linewidth]{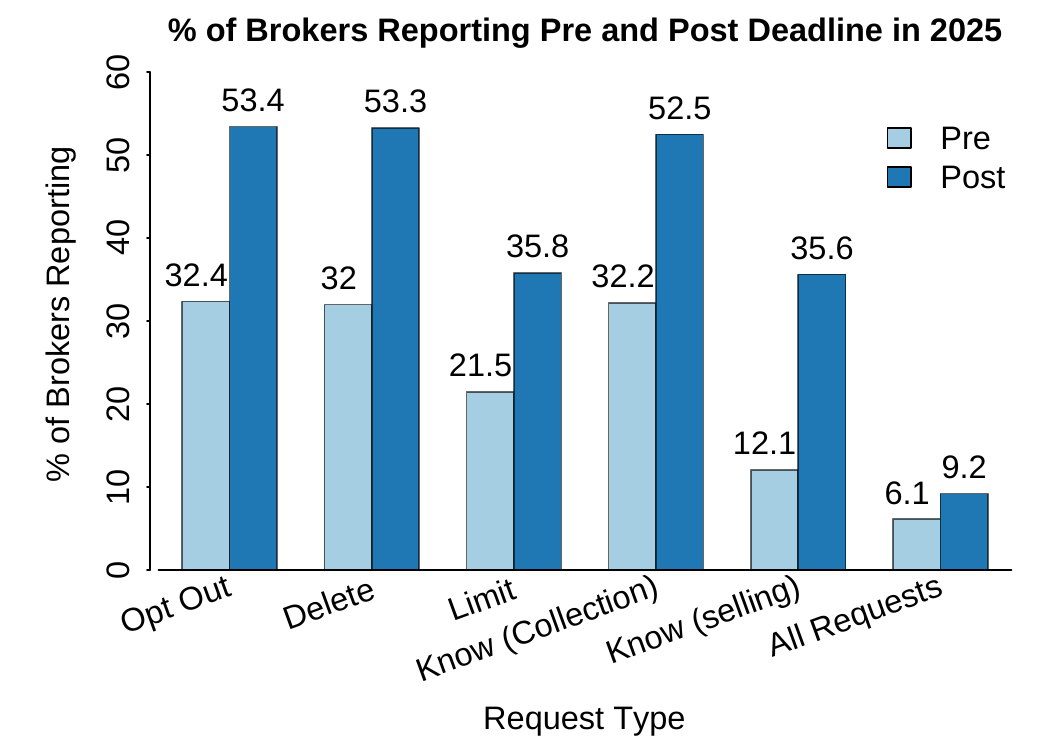}
    \caption{Percentage of data brokers reporting all request types total and every specific request type before and after the July 1st, 2025 regulatory deadline. The deadline increased compliance with the transparency requirement, as demonstrated by the increase in post-deadline reports, though reports remain under 55\% for any type of request}
    \Description{This figure illustrates key ``Pre'' and ``Post'' regulatory deadline rights reports.}
    \label{fig:percent-reporting-pre-post-deadline}
  \end{minipage}
  \hfill
  \begin{minipage}[t]{0.48\columnwidth}
  \vspace{0pt}
    \centering
    \includegraphics[width=\linewidth]{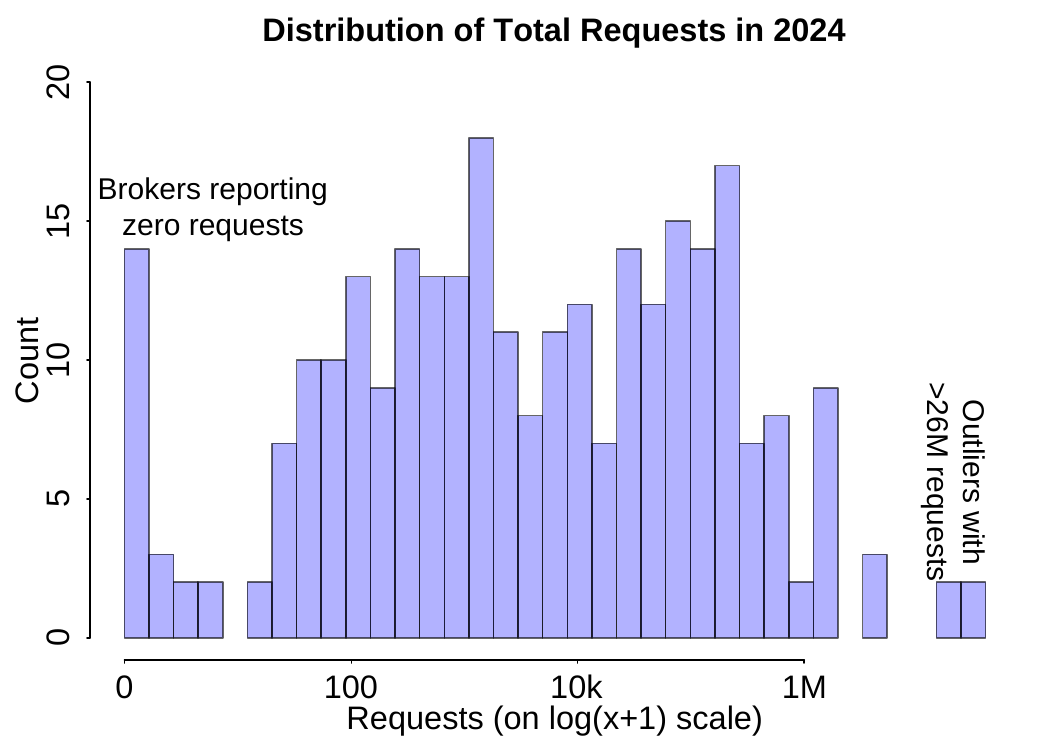}
     \caption{Distribution of the total requests received in 2024, including brokers that explicitly report zero requests. The majority of brokers that report any requests received fall under a total of 50,000 requests, while 17 brokers received requests in the millions.}
     \Description{This figure shows a right-skewed distribution of the total number of requests received by brokers in 2024}
    \label{fig:distribution-total-requests}
  \end{minipage}

\end{figure}

\textbf{Noncompliance with Transparency Requirements.} Of the 522 data brokers registered in California, we found that 45\% do not report any requests across all rights categories. After the compliance deadline of July 1st, 2025, only 9.2\% were fully compliant with \textit{all} transparency requirements, as seen in Figure \ref{fig:percent-reporting-pre-post-deadline}. The request types with the highest numbers reported are opt-out of sales and deletion: 53\% reported receiving do not sell requests, and 53\% reported receiving deletion requests, which also comports with data reported by CPPA \cite{cppa-enforcement-update-2025}. Figure \ref{fig:percent-reporting-pre-post-deadline} also shows the change in requests reported before the regulatory deadline of July 1st, 2025 (light blue) and 45+ days after the deadline (dark blue). There was a significant increase in the percentage of reports per request type after the deadline ($p$-value = 0.02). Brokers that report requests received also report the number of requests fulfilled, with fulfillment rates over 80\% (except for requests to know, which has a 73\% fulfillment rate). %Table \ref{table: inconsistencies-in-reporting} shows that even among reporting brokers, many exhibited inconsistencies in their reporting, such as an unverifiable reporting period. 

Figure \ref{fig:distribution-total-requests} illustrates the distribution of total requests received in 2024. Variation is large: the 25th percentile of total requests received by brokers was 221;  the median was 3,268; and the 75th percentile was 63,170. The broker with the highest number of requests reported 29,627,364—over nine thousand times the median. %However, these distributions appear to not be unique to data brokers. 
To put these totals in perspective, we collected request report numbers for a random selection of 20 businesses (non-data brokers) determined by Tran et al. (2025) to be within the top 1000 visited websites subject to CCPA compliance. While the median total requests is higher (20,733), many of these companies similarly do not report all rights requests and some receive fewer than 1,000 total requests. %We discuss our exploratory investigation of correlates with the number of requests in Section 4.3. 
Outliers may be at least partially explained by the use of third party services, such as Consumer Report’s Permission Slip, DeleteMe, Optery, and similar businesses that submit opt-out (do not sell) and deletion requests to some brokers on behalf of consumers. Unfortunately, most of these services do not publish complete lists of the companies for which they submit requests, making this difficult to verify. %so we were unable to verify across multiple services which data brokers from the registry were the recipients of requests from these third party services. 

Even among reporting brokers, transparency remains inconsistent. 44\% of brokers published metrics whose reporting year was unverifiable or incorrect, either because the date the privacy policy was updated was unclear, predated the 2024 reporting period, or explicitly noted that the metrics were for an incorrect year; 15\% of brokers report total request numbers but omit how many requests they complied with, making it difficult to assess fulfillment rates; 7\% of brokers report total request numbers that do not match the sum of “complied with” and “denied” requests, calling into question the validity of the numbers; and 1\% of brokers lacked a privacy policy altogether and thus didn't post any metrics. 

%\begin{table}[t]
%\centering
%\caption{Inconsistencies and gaps in publicly posted CCPA request metrics.}
%\label{fig:inconsistent-metrics}
%\begin{tabular}{p{3.8cm} p{8.2cm} r}
%\toprule
%Issue observed & Description & \% of brokers \\
%\midrule
%Unverifiable or incorrect reporting period &
%Reporting year unverifiable or incorrect (e.g., policy update date unclear, predates the 2024 reporting period, or explicitly reports metrics for a different year) &
%44 \\
%Incomplete compliance disclosure &
%Reports total requests but omits how many requests were complied with &
%15 \\
%Internal inconsistency &
%Reported total requests do not equal the sum of ``complied with'' and ``denied'' requests &
%7 \\
%Missing policy &
%No privacy policy found &
%1 \\
%\bottomrule
%\end{tabular}
%\end{table}

%\begin{figure}[t!]
    %\centerline{\includesvg[width=0.75\columnwidth]{figures/fig1.svg}}
   % \caption{Example of using SVG on Overleaf}
    %\label{fig: percent-reporting-pre-post-deadline}
    %need to put in describtion but im confused where?
%\end{figure}

% ADD IN APOORVAS TABLE HERE WITH LABEL distribution-total-requests

\textbf{Challenges with Exercising Privacy Rights.} Table \ref{tab:rq2-results}  presents the prevalence of all design features that either demonstrate noncompliance with the CCPA or increased friction in the submission process across our sample of 250 data brokers. When reporting percentages of brokers that use noncompliant design features, we weight the results from the stratified sample of 250 brokers to be representative of the full population of 522 registered brokers. First, we find that 43\% of data brokers do not allow consumers to exercise all six of their rights. We observed that brokers primarily enable consumers to exercise their right to “do not sell” my data (84\% of brokers), delete their data (88\%), and to know what data is collected (92\%). 37\% of data brokers require consumers to verify their identity to exercise their do not sell right or to limit the use of their sensitive personal data, which is explicitly forbidden under the CCPA (Cal. Code Regs. § 7026(d)). Second, we find that 64\% of brokers have at least one design feature in the rights request process that increases submission friction. 43\% of data brokers require multiple submissions to fulfill each category of request, even if the information required to submit each request is identical. %Figure \ref{fig: example-of-multiple-forms} provides an example of a form that would allow consumers to submit all requests in one versus a form that requires multiple submissions of the exact same form for each right. 
22\% of brokers require a CAPTCHA submission in request processes, an extraneous feature found in previous work to be a barrier to exercising privacy rights given the unlikelihood of brokers receiving fraudulent requests \cite{tran-2025}. In an extreme case, one request to opt out required solving eight CAPTCHAs in a row, including math equations and identifying differences in rotation of objects. Third, after submitting a request, the process can further induce confusion. Submission confirmations are inconsistent and, in some instances, impossible to attribute to a specific broker; some brokers post a confirmation page, others send a confirmation email, but others provide no confirmation at all, or send an email confirmation or a request for further verification under a different company name. For example, one author received an email from an unknown company to verify their first and last name to complete their rights request (without stating which type of request was submitted), even though this company name was not associated with any of the data brokers to whom we had submitted requests.
\begin{table}[]
\caption{Weighted percentage of brokers registered in 2025 that utilize noncompliant or friction design features in the request submission process}
\label{tab:rq2-results}
\resizebox{\columnwidth}{!}{%
\begin{tabular}{@{}llll@{}}
    \toprule
    Type               & Feature                                                                                               & Description                                                                                                                                                                                                             & \% of Brokers \\ \midrule
    Noncompliance     & \begin{tabular}[c]{@{}l@{}}Not all requests\\ available\end{tabular}                                  & \begin{tabular}[c]{@{}l@{}}Broker does not provide a process to submit all privacy\\ rights requests (Cal. Civ. Code § 1798.105 to § 1798.121)\end{tabular}                                                                & 43\%          \\
                       & \begin{tabular}[c]{@{}l@{}}Missing privacy\\ policy\end{tabular}                                      & \begin{tabular}[c]{@{}l@{}}Broker website is missing a privacy policy \\ (Cal. Civ. Code § 1798.99.85)\end{tabular}                                                                                                         & 3\%           \\
                       & \begin{tabular}[c]{@{}l@{}}Excessive \\ information\end{tabular}                                      & \begin{tabular}[c]{@{}l@{}}Broker requires searching or scrolling through information \\ after clicking a submission request \\ (Cal. Code Regs. Tit. 11, § 7004(a))\end{tabular}                                         & 30\%          \\
                       & \begin{tabular}[c]{@{}l@{}}Verification required\\ where unnecessary\end{tabular}                     & \begin{tabular}[c]{@{}l@{}}Broker requires identify verification for requests that should not\\ require it (request to opt out (Cal. Code Regs. § 7026(d)),\\ or request to limit (Cal. Code Regs.§ 7027(e))\end{tabular} & 37\%          \\
                       & Broken link(s)                                                                                        & \begin{tabular}[c]{@{}l@{}}Broker has invalid links to submissions and/or invalid email\\ address or phone number (Cal. Code Regs. Tit. 11, § 7004(a))\end{tabular}                                                       & 10\%          \\ \midrule
    Increased Friction & CAPTCHA test                                                                                          & \begin{tabular}[c]{@{}l@{}}Broker requires solving a CAPTCHA test in any of the\\ submissions \end{tabular}                                                                                     & 21\%          \\
                       & \begin{tabular}[c]{@{}l@{}}Separate / \\ multiple forms\end{tabular}                                  & \begin{tabular}[c]{@{}l@{}}Broker requires that a consumer resubmit the same form \\ multiple times or submit multiple forms, despite collecting \\ the same information\end{tabular}                                   & 43\%          \\
                       & \begin{tabular}[c]{@{}l@{}}Difficult to access or \\ sensitive info required\\ to submit\end{tabular} & \begin{tabular}[c]{@{}l@{}}Broker asks for information in excess of email, name, address,\\ or telephone, such as: Device ID, advertising ID, \\ Social Security Number or other personal information\end{tabular}& 20\%          \\ \bottomrule
    \end{tabular}%
    }
\end{table}

\textbf{Correlates of Request Volumes.} We analyzed the differences between brokers that did not report any metrics (and are hence plainly in violation of the Delete Act) and those that did. Conducting an ANOVA on the logistic regression results, we find that the only statistically significant predictor is corporate subsidiary status, with brokers that are subsidiaries being more likely to report rights requests. An analysis of variance indicated that brokers that are subsidiaries of larger companies are significantly associated with reporting ($p$-value = 0.046; results from this logistic regression are reported in Appendix \ref{logistic-regression-appendix} Table \ref{tab:coefficients-from-regressions}). While not a significant difference, we find that data brokers that do not report have, on average, a smaller annual income (over \$94M vs. over \$402M for those that report requests) and fewer employees (391 vs. 1,661 for those that report) than those that do report any request metrics. 

%94,198,082 US dollars vs 402,011,275 US dollars for those that report

Second, in investigating the variation in total requests received, we found neither substantial (or statistically significant) differences in the data broker self-reported features nor in the corporate entity features. While the differences are not significant they are observable. For example, data brokers that receive more requests have, on average, higher annual income and more employees (as seen in Table \ref{tab:brokers-reporting-differences}). We conduct the same analysis per request type, with observable differences mainly associated with the type of data a broker collects, whether or not the broker is a subsidiary, and if the broker is subject to other laws. Tables \ref{tab:right-to-delete-differences} through \ref{tab:right-to-know-differences} in Appendix A.4 show all feature differences for each request type with measures of significance. We similarly investigated the sample of 250 brokers we reviewed for friction in their consumer request processes and found that brokers that received fewer requests are more likely to not allow consumers to exercise all required privacy rights (p-value = 0.0001). Similarly, brokers that received fewer requests also have a significantly higher proportion of excessive or confusing information in their request processes (p-value = 0.02). Table \ref{tab:friction_to_requests} reports the differences across the groups of lower request receiving and higher request receiving brokers for all friction features we measured.

%finding that data brokers that are subject to FCRA receive significantly more requests to limit the sharing of personal data than those that are not subject (Table \ref{tab:right-to-limit-differences} in Appendix A.4). We also find that data brokers that are subsidiaries receive significantly more requests to know what is being collected / shared (Table \ref{tab:right-to-know-differences}). 

\begin{table}[]
\caption{ Comparison of correlates between data brokers receiving low and high total numbers of requests. *Denotes a subgroup of data brokers with lower (N=65) and higher (N=66) request amounts is due to unavailable corporate entity data for all data brokers. **P-values for proportion of data brokers with a certain characteristic is calculated using a two sample proportion test while p-values for corporate income and number of employees are calculated using t-tests.}
\label{tab:brokers-reporting-differences}
\resizebox{\columnwidth}{!}{%
\begin{tabular}{@{}llllll@{}}
\toprule
                                                                                  & \begin{tabular}[c]{@{}l@{}}Low requests received \\ (N=142)\end{tabular} &            & \begin{tabular}[c]{@{}l@{}}High requests received\\ (N=142)\end{tabular} &             &         \\ \midrule
                                                                                  & Percentage                                                               & SE         & Percentage                                                               & SE          & p-value** \\ \cmidrule(l){2-6} 
\begin{tabular}[c]{@{}l@{}}Collects data from minors (\%)\end{tabular}          & 2.1                                                                      & 1.2        & 5.6                                                                      & 1.9         & 0.21    \\
\begin{tabular}[c]{@{}l@{}}Collects precise geolocation data  (\%)\end{tabular} & 18.3                                                                     & 3.2        & 14.1                                                                     & 2.9         & 0.42    \\
\begin{tabular}[c]{@{}l@{}}Collects reproductive health data (\%)\end{tabular} & 1.4                                                                      & 1          & 1.4                                                                      & 1           & 1       \\
Subject to FCRA (\%)                                                              & 1.4                                                                      & 1          & 4.2                                                                      & 1.7         & 0.28    \\
Subject to GLBA (\%)                                                              & 3.5                                                                      & 1.5        & 5.6                                                                      & 2           & 0.57    \\
Subject to CMIA (\%)                                                              & 1.4                                                                      & 1          & 0                                                                        & 0           & 0.48    \\
Subject to IIPPA (\%)                                                             & 0.7                                                                      & 0.7        & 0                                                                        & 0           & 1       \\
Subject to HIPAA (\%)                                                             & 3.5                                                                      & 1.5        & 7                                                                        & 2.1         & 0.29    \\
From California (\%)                                                              & 21                                                                       & 0.04       & 23.2                                                                     & 3.5         & 0.89    \\
Is subsidiary*                                                                    & 25                                                                       & 5.5        & 27                                                                       & 5.5         & 1       \\ \midrule
                                                                                  & Mean                                                                     & SE         & Mean                                                                     & SE          & p-value** \\ \cmidrule(l){2-6} 
Income*                                                                           & 114,177,857                                                              & 42,158,682 & 685,347,296                                                              & 475,529,982 & 0.24    \\
Employees*                                                                        & 641                                                                      & 276        & 2,666                                                                    & 1,939       & 0.31    \\ \bottomrule
\end{tabular}%
}
\end{table}

\begin{table}[t]
\centering
\caption{Difference in means of identified noncompliance or friction features by the volume of requests received, where high and low reporting is split based on median request value.}
\label{tab:friction_to_requests}
    \begin{tabular}{lrrrrr}
    \toprule
     & \multicolumn{2}{c}{Lower Reporting (N = 125)} & \multicolumn{2}{c}{Higher Reporting (N = 125)} &  \\
    \cmidrule(lr){2-3} \cmidrule(lr){4-5}
     & Mean & SE & Mean & SE & p-value \\
    \midrule
    Average number of features identified
     & 2.14 & 0.11 & 1.82 & 0.13 & 0.06 \\
    \textbf{Not all requests available (\%)}
     & \textbf{55} & \textbf{5.00} & \textbf{18} & \textbf{3.84} & \textbf{$<$0.001} \\
    \textbf{Excessive or confusing information (\%)}
     & \textbf{29} & \textbf{4.50} & \textbf{15} & \textbf{3.57} & \textbf{0.02} \\
    Verification required (\%)
     & 40 & 4.90 & 46 & 4.90 & 0.47 \\
    \textbf{CAPTCHA test (\%)}
     & \textbf{14} & \textbf{3.50} & \textbf{33} & \textbf{4.70} & \textbf{0.003} \\
    Separate forms (\%)
     & 48 & 5.00 & 36 & 4.80 & 0.12 \\
    Difficult to access or sensitive information (\%)
     & 15 & 3.60 & 24 & 4.27 & 0.15 \\
    \bottomrule
    \end{tabular}
\end{table}

Third, we investigated the role of third party submission tools. We were able to identify brokers that received requests from ``WebChoices'', an online tool from the Digital Advertising Alliance, an industry trade group. WebChoices allows consumers to submit do not sell requests to 97 companies from a single interface. In this list, we identified 48 registered data brokers in their tool and analyzed whether these participating brokers received more do not sell requests than registered brokers not featured in WebChoices. Brokers that are featured in WebChoices receive 58\% more requests to opt out, on average (633,286 vs 399,847 requests). 

In order to compare and distill differences in data broker non-compliance, we created two ranking indexes based on the variables we identified in our analysis. The first scores brokers based on the amount of friction in consumer requests, categories of sensitive data collected, and whether they comply with transparency requirements. Because this uses data from our friction analysis that was based on a stratified sample of 250 brokers, we also created a second index that ranks based on transparency and sensitive data categories across the entire set of brokers. Both rankings produce opacity scores, where a higher opacity score represents less compliance to the CCPA. Our indexing framework and results is located in Appendix A.7.

\section{Limitations}

Our work has several limitations. First, our assessment of compliance was performed at two time periods---once in the months before the effective date of July 2025 and once 45 days after. The surprisingly low compliance rate may be an artifact of data brokers still working on reporting these metrics in their privacy policies. The evidence shows a statistically significant increase between the two periods. 
That said, data brokers were on notice for over three years, and compliance even after reporting became mandatory remains surprisingly low. We also experimented with developing an automated solution, using a combination of webscraping and large language models to classify compliance, but found that the heterogeneity of privacy policies made that challenging. Below we discuss the more ideal solution, which is more unified reporting. 

Second, some might argue that our overall compliance measure of 9\% is stringent. A data broker must report requests received for all five request types, and thus we also report each distinct request type separately which exhibit higher individual compliance rates. Even then, compliance hovers at just above 50\% at best. But full compliance remains a relevant measure: brokers do not have the discretion to comply with only a selected subset of transparency requirements. %that report some, but not all, requests received begs the question of whether the  broker allows consumers to exercise only those reported rights requests; an important consideration for our investigation into a consumer's ability to exercise all their rights. 

Third, our compliance assessment focuses on transparency, and perhaps the other components of California's privacy laws that we do not measure here are more substantively important.  We do, however, assess consumer friction and other potential violations of the request process, and the difficulties in implementing the transparency provisions undercut the public intention to enable researchers, public interest groups, and consumers to understand how data brokers are acting. 

Fourth, some of the consumer frictions we document represent a tradeoff between verification security and ease of the consumer process, though the CCPA stipulates that businesses may not require identity verification for the do not sell and limit rights. We do, however, document apparent violations of California law and our audit of these processes suggests that verification may often stand as a shield against consumers exercising their rights. 

Fifth, our analysis of the correlates of reporting volume are descriptive and not causal in nature.\footnote{In addition, these tests may be sensitive to outliers, which is why we focus one inquiry on reporting vs. non-reporting brokers.} They hence only provide suggestive evidence as to differences in types of data brokers. Moreover, the difficulties in matching subsets of data brokers to auxiliary data mean that even these descriptive inferences may not generalize to the population of data brokers. 

Last, our analysis documents compliance with then-current California law. In 2026, California is introducing a new mechanism (the Delete Request and Opt-Out Platform or DROP mechanism), which purports to be a centralized, one-stop shop for ``dropping'' data from all registered data brokers. While this mechanism is promising, as we discuss below, our work highlights how important it will be to monitor the effectiveness of this mechanism. 

\section{Discussion}

In spite of the fanfare surrounding California's consumer privacy laws, our findings paint a troubling picture of how these protections have been implemented. The majority of data brokers do not comply with California's transparency requirements. And consumers face considerable confusion, obfuscation, and frustration should they try to exercise their statutory rights. These findings and California's attempts to reach data brokers have important implications for the privacy landscape, particularly as regulators elsewhere have struggled to apply general-purpose privacy frameworks such as the GDPR to opaque data broker markets.

\textbf{The Importance of Registration.} This assessment was enabled by the fact that data brokers are required to register with the state of California. Similar evaluations of the CCPA on non-data brokers have struggled to verify which businesses are covered by the law given the eligibility requirements and a corresponding lack of registry to identify them \cite{tran-2024}. To be sure, the Data Broker Registry features a \textit{self-selected} group of brokers that elected to register, meaning at minimum these businesses were concerned enough to at least register with the CPPA. There are penalties for failing to register, and the CPPA has engaged in sweeps to identify non-registered brokers doing business in California \cite{ccpa-fines-2026}. Yet even for the brokers that do register we observed weak compliance with statutory transparency requirements, though the July 2025 regulatory deadline for posting request metrics on their websites did increase reporting as compared to the metrics reported in the registry. Ultimately, only 55\% of registered data brokers report any type of request metric on their website post-July 2025. The value of this registry, however, raises the question why \textit{all} businesses subject to the CCPA are not similarly required to register to declare that they are subject to the law. New York's Local Law 144, which was widely heralded as an exemplar for mandating third party audits of algorithmic tools in employment, exemplifies this weakness of what Wright et al. characterized as ``null compliance,'' making it impossible for the public or regulators to determine who is subject to the law \cite{wright-2024}. In a recent audit report, the New York Comptroller General confirmed that enforcement has been, as a result, lackluster \cite{dinapoli-2025}. Knowing \emph{who} is subject to the law is a critical first step. 

\textbf{The Legislative Compromise around Public Enforcement.} Our findings show that California's data privacy law, while strong on paper, appears weak in practice, at least in the dimensions we assessed. Worth understanding here is the political economy context. Policymakers found themselves bargaining with the tech industry in 2018 to avoid the certainty of the CCPA passing as a state ballot initiative with the private right to action intact, instead brokering a last minute compromise to remove the CCPA from the ballot and pass it through the legislature \cite{confessore-2018}. This resulted in a key compromise around enforcement: namely the lack of a general private cause of action that would have enabled individuals to sue businesses directly for violations \cite{iapp-proa}. (The CCPA private right of action is limited only with respect to security breaches and the Delete Act does not provide a stand alone right of action against brokers.) Instead, Californians must submit suspected violations to the CPPA to trigger public enforcement. 

The same issue divides policymakers in Washington as two versions of federal privacy laws have been proposed but not advanced since 2022, with a private right of action being one of the key issues that stymied agreement \cite{iapp-proa, johnson-2024-itif}.\footnote{The other is preemption of state laws. An April 2026 legislative proposal by House Republicans both preempts state privacy laws and does not include a private right to action.} While the CPPA's enforcement work as a young agency is commendable and growing, it is also limited by the reality of scarce resources. Given the substantial evidence of underenforcement, there are strong reasons to update the CCPA to include a private right of action and allow aggregation of claims against data brokers for violations of the law.

The penalty structure further explains noncompliance with transparency requirements. While data brokers are subject to fines of \$200 per day for failing to register or for not deleting data on time, there are no fines for failing to publicly post rights request metrics. The lack of penalties undermines the check on noncompliance that public posting is intended to aid. Previous work investigating compliance to laws that require an actor to provide transparency have shown similar patterns of lack of compliance \cite{crain-2017, urban-2020, wright-2024}. Our findings suggest that penalties should attach to the failure to comply with transparency requirements, such as not posting metrics, using dark patterns in the request process, and failing to have a working privacy policy page.

\textbf{Compliance Delegation to Regulated Parties.} Another structural trend exemplified by the CCPA is that it delegates key compliance decisions -- even registration -- to regulated parties. It took substantial efforts to manually collect the information in this study from a cacophony of privacy policies. Policymakers aimed to have academics, civil society, and others crowdsource compliance, but the reality is that this delegation decision makes systematic monitoring costly, even with a known, identified population of registered data brokers. There is a certain irony that data brokers---who promise centralized records---evade monitoring by decentralized reporting. The obvious solution here would be to require a common, standardized repository of all compliance information by brokers. For example, requiring brokers to report in both human and machine-readable formats (such as a standardized JSON or XML format for privacy metrics), to separate Californians from non-Californians, and to denote requests received via individual request processes versus the DROP mechanism. Because brokers already collect geographic indicators to determine whether or not they are required to comply with a given request, reporting geographic information should not impose substantial new burdens.

\textbf{Increasing the Clarity and Consistency of Reported Metrics.} Evaluating the effectiveness of the law’s impact on increasing the transparency of data brokers’ business practices was exceptionally challenging due to the ambiguities of the reporting requirements.
%While we hypothesized that the registry metrics would overlap with the website posted metrics, this was not the case. 
Brokers are required to report request metrics for \textit{two years} prior to the year they register (Cal. Civ. Code § 1798.99.82)\footnote{The 2025 registry, for instance, contained request metrics for 2023.} But confusingly, the Delete Act's public posting requirement is for the \textit{previous calendar year,} meaning that the data posted on July 1, 2025 should include January to December 2024. This gap is confusing and requires clearer explanation. %We elected to manually review all broker privacy policies both pre- and post- July 2025 to collect these metrics, a labor-intensive task. 

Additional analysis we conducted suggests that the majority of brokers posted metrics that were at best incomplete, and in some cases, likely incorrect. After the completion of our study, the 2026 version of the registry was released in April (compiling 2024 metrics). As a robustness check we compared these requests to our own manually collected requests. The registry reported metrics should have, in theory, been identical. However, this was not the case: of the 457 brokers we were able to match across datasets, only 27 brokers (5.9\%) had metrics that matched identically, only 45 brokers (9.8\%) reported the same metric coverage across both sources, and 77 (17\%) reported metrics to the registry that were \textit{lower} than their posted metrics. In addition, for 43.5\% of matched brokers, the 2026 registry contained a value for a metric that we did not observe in our hand-coded dataset, meaning that brokers were not fully reporting their 2024 metrics by the July 1st deadline. We discuss these findings in greater depth in Appendix~\ref{2026-registry-appendix}.
These findings suggest that centralizing disclosures should make reported metrics more consistent, accessible, and easier to monitor for compliance, but clearer definitions and machine-readable reporting standards remain necessary. But the Delete Act's public posting requirement is especially flawed given that brokers appear to have little incentive to comply with it, and that verifying each set of metrics is today is such a challenge.  Finally, using the newly submitted registry data, we analyzed whether there were differences in self-reported variables such as sensitive data types collected and applicable laws, as the 2026 registry included the same questions (as well as several new data types). While this analysis is not directly relevant to our study, we include results in Table \ref{tab:2026-bivariate-analysis} for readers to see the differences between low and high reporting data brokers based on the metrics they report in the 2026 registry. 

%Ultimately the disjoint between both metrics reporting requirements creates confusion, potentially for both brokers and the public.

\textbf{Frictions in Request Processes.} The opt-out framework of the CCPA places the burden of exercising privacy rights on individual Californians, which means that the design and usability of the request processes are crucial to their efficacy. However, we found that many data brokers do not allow consumers to exercise all of their privacy rights and often use design features in the submission process to make it harder for consumers to exercise them. %We found that 43\% of our sampled data brokers only allow consumers to exercise their privacy rights by submitting multiple forms that require the same information, making the process more laborious and time consuming. We also identified design features that are not compliant with the Delete Act, which requires brokers to limit the use of dark patterns that can cause friction in the submissions process (CA Civ. Code §1798.99.82). %Given the time demands of submitting requests, multiplied by the number of registered data brokers and the additional friction added by some to the request processes, the burden for consumers to exercise their data privacy rights is untenable.
Fortunately, California is shifting to a more systemic approach for automating the exercising of privacy rights with the new DROP mechanism, effective January 1, 2026. Brokers are required to process consumer DROP requests starting August 2026 and update them every 45 days. But the success of DROP in reaching the California consumers who want to exercise do not sell and deletion rights with data brokers will hinge on how effectively the CPPA and consumer rights organizations conduct outreach to publicize DROP’s existence. Given California’s large population, even modest adoption among Californians will have an impact on the data marketplace. And importantly, a demonstrated appetite for and willingness to enroll in DROP provides support for the adoption of opt-out preference signals (or OOPS, also known as Global Privacy Control), which browser developers are required to support by January 2027. The shift towards automating rights is a significant improvement, though auditing compliance will remain a challenge. %OOPS were introduced in the CCPA as an optional method for exercising do not sell requests and adopted only by the Mozilla Firefox and Brave browsers. But as of 2026 per the Opt Me Out Act (AB 566) as of January 2027 California will be the first state requiring all browser developers to allow consumers to set opt-out of sale (do not sell) signals to automate these processes.

%Again, given that consumers must either sign up for DROP or choose to implement OOPS in their browsers, the challenge for adoption lies with informing consumers of their existence and ensuring that enabling these mechanisms is simple. 

Beyond the automating of rights requests, identifying and regulating friction in the request process is still an important issue for California \cite{calprivacy-2026}. Design matters. When Apple first launched a feature for consumers to limit tracking by advertisers on the iPhone, adoption of this privacy feature was low because the option was buried in the iPhone’s settings. Once the company placed the feature (renamed App Tracking Transparency, or ATT) front and center for consumers as a dialog window that appears when first opening an app in iOS 14 (2021), adoption of ATT soared \cite{apple-2025}. When privacy options are made visible and simple for consumers to adopt, they tend exercise them overwhelmingly, matching decades of public opinion research of consumers stating their preferences for data privacy and disputing the so-called privacy paradox argument that consumers say they care about privacy but reveal their true preferences by continuing to use privacy-invasive technologies. Ultimately, where consumers are forced to do the piecemeal work of exercising rights, companies run interference to make the process difficult. 

CPPA recently initiated an enforcement action based on the use of design friction to interfere with consumer rights requests. In March 2026, the CPPA board fined Ford Motor Company \$375,703 for requiring consumers to verify their identity when submitting do not sell requests (which per statute cannot require verification) \cite{ford-2026}. Reviewing request processes for friction-based indicators could aid regulators with prioritizing enforcement actions, particularly during the transition period before automated mechanisms are operational. However, our identification of high rates of non-compliance and friction in brokers' request processes suggests a reconsideration of how request processes should be managed. Currently, any business subject to the CCPA can use any design they wish as long as it complies with statutory requirements, such as prohibiting dark patterns. We argue that a more top-down approach, such as mandating a set of templates that businesses must adopt to make the process uniform, will  reduce friction by eliminating ambiguity and designs that creatively increase friction while skirting the letter of the law.

%While the Delete Act will require brokers to conduct independent audits of their compliance with registry reporting requirements every three years beginning in 2028, the law does not attach fines for failing to conduct such audits nor require that the results be routinely submitted to the CPPA. As a result, comprehensive oversight of broker practices may remain difficult in practice, especially because the human resources and capacity to conduct such a review may be limited. 

%While the Delete Act will require brokers to conduct independent audits of their compliance with the reporting requirements every three years starting in 2028, the law doesn't attach fines for non-compliance with the audit requirement. Nor does it require brokers to submit their audit results to the Agency, but only to have them available if asked. These rules continue to give brokers the option of making a bet that evading the rules will be a better payoff than the risk of enforcement.

\textbf{Population Coverage.} The CCPA allows businesses to include non-Californians’ requests in their rights reporting, making it challenging to isolate the impact of California’s laws on Californians. While businesses must state whether they are including non-Californians in their metrics, they do not have to disclose which requests originate from California (Cal. Civ. Code § 1798.99.85). In our review of policies, we find that 22\% explicitly state they are reporting California requests, 9\% report they are U.S. or global, while 69\% make no mention from where location requests originate. While this choice was made to ease compliance burdens on businesses who do not verify residency, this mixing of data is problematic; if Californians are to understand if the CCPA is working \textit{for them, }it must be clear how many of \textit{their} requests are being fulfilled. Requiring companies to track California-specific requests separately would provide more granularity into exactly how the CCPA is helping California consumers; the fact that only Californians can submit DROP requests may address this problem after 2026.

\textbf{The Puzzle of Corporate Structure.}
%Despite the aims of the CCPA and Delete Act, not only did we find evidence of noncompliance with the law, we also observed evidence that suggests that brokers are attempting to evade registry through corporate structures, are failing to report all their subsidiaries, and reporting identical metrics across subsidiaries.
In our efforts to understand what influences which brokers do and do not comply with transparency requirements, we found that whether a broker reports metrics at all is significantly associated with whether the data broker is a subsidiary of another company. %For example, across all request types, data brokers that are known subsidiaries received, on average, 1,530,504 requests in 2024 while parent companies received an average of 174,077. One interpretation of this behavior is that maintaining multiple subsidiaries provides some insulation from consumer rights requests as it increases the number of requests consumers must make. While it was difficult in many cases to clearly trace relationships between companies, we observed several behaviors that suggest relationships. For example, multiple independently registered brokers appear to be part of the same company but under different names, and some even reported identical metrics numbers across companies. We refer to such data brokers as “silent duplicates”, a term coined by Gerchick et al.  \textbackslash{}cite\{gerchick-2025\}. 
On December 17th, 2025 the CPPA released an advisory, noting that some data brokers “may be making it difficult for consumers to identify them by using trade names or websites that do not appear on their annual registration” \cite{ccpa-advisory-2025}. Our research similarly revealed that some brokers were engaging in practices that appeared to obscure corporate relationships. For example, we identified silent duplicates \cite{gerchick-2025}, with nearly or completely identical privacy policies, but by entities that did not reference one another or list identify them as a parent company or subsidiary (see Appendix \ref{sec:silentdup}). We found examples of silent duplicates that have identical metrics, privacy policy links, or contact information, yet are based in different states. While ``borrowing'' of legal language may be common, identical policies within the same industry of seemingly unrelated parties may be indicative of a broker using multiple entities to increase the barriers for data deletion. %Our evidence supports this assertion, though it is possible brokers are also trying to evade the attention of the CPPA, as well as consumers, through complex relationships and evading registration.

These findings suggest that more detailed documentation of corporate structure in the data broker registry would help both to empower consumers and  better understand how brokers share information and honor consumer requests between companies. If the intent is to crowdsource compliance, such corporate structure information would allow parties to assess how different request channels influence the number of consumer requests and denial rates, which is currently obscured information due to the heterogeneity of broker reporting.

\section{Conclusion}
Data brokers have been understudied in the context of compliance with privacy laws in the United States. Our research demonstrates how these companies fail to meet transparency and reporting requirements as set forth by the CCPA and the Delete Act. For those brokers that do comply, we find that they have high fulfillment of consumer privacy requests under the CCPA. Despite this, our qualitative review of 250 data brokers’ request processes finds that over 43\% do not allow consumers to exercise all of their privacy rights. Finally, we find large discrepancies in the number of requests received, with the majority of data brokers receiving fewer than 60,000 total requests and outliers receiving over 20 million. We observe these discrepancies can be due to how large the data broker is, if the broker is a subsidiary and respects requests submitted to parent companies, and the ease with which a consumer can exercise a request via the data broker. We provide evidence supporting the California Privacy Protection Agency’s development of a streamlined process to opt out consumers out of the selling of their data and delete and argue for streamlining all other request processes, as well. Additionally, we argue that to increase compliance with transparency requirements, the CPPA should release a standardized format that data brokers must follow to report the necessary requests received and complied with, and from which locations, in order to enhance quality of the data meant to increase transparency. Similarly, uniform design templates for consumer request processes can help decrease friction in how brokers' allow consumers to exercise their privacy rights. We present the first analysis of data broker compliance with the Delete Act and the CCPA for all privacy requests, and suggest future work that continues to focus on all privacy rights and enhances the infrastructures for enabling consumers to exercise their rights and agency over their data.

\begin{acks}
We are grateful to Caroline Yee and Jason Shin for their research assistance with this project, and the summer 2025 Stanford RegLab cohort for their suggestions.
\end{acks}

\section*{Generative AI Usage Statement}
All authors certify that they did not use any large language model or language generation tool in writing this publication. ChatGPT and Claude were used to find code \textit{examples} of restructuring table formats to booktabs, and that code was adapted and verified before it was inputted into this manuscript. 

\bibliographystyle{ACM-Reference-Format}
\bibliography{references}

\newpage
\appendix

\section{Appendix}
\subsection{Codebook for Evaluating CCPA and Delete Act Compliance}

In order to answer the questions of 1) how many data brokers comply with the transparency requirements to post requests received from consumers and 2) what are the requests received from consumers in 2024 to data brokers, two authors manually reviewed the 522 privacy policies of all registered data brokers in California. In order to do so rigorously and reproducibly, they iteratively developed a codebook based on requirements of the California Civil Code § 1798.99.85 (the Delete Act's transparency requirements). The final codebook used to gather data from 522 privacy policies before July 1st, 2025 and the same 522 privacy policies after July 1st, 2025 is detailed in Table \ref{tab:codebook-rq1-collection}. 

\begin{table}[h]

\caption{Codebook followed by authors to collect data on the request metrics reported on data broker privacy policies}
\label{tab:codebook-rq1-collection}
\resizebox{\columnwidth}{!}{%
\begin{tabular}{@{}llll@{}}
\toprule
Data to be collected                                                                                                & Value type              & Description                                                                                                                                                                                                                                                                                                                                                                                                                                                                                 &  \\ \midrule
Row \# in 2025 registry                                                                                             & Integer                 & The original row number the data broker is associated with in the 2025 registry                                                                                                                                                                                                                                                                                                                                                                                                             &  \\
Broker                                                                                                              & String                  & Name of data broker as it appears in the 2025 registry                                                                                                                                                                                                                                                                                                                                                                                                                                      &  \\
Link to privacy policy                                                                                              & URL or none             & Working privacy policy for data broker, none if no privacy policy given in registry / found                                                                                                                                                                                                                                                                                                                                                                                                 &  \\
Link to metrics                                                                                                     & URL or none             & URL with where California consumer request metrics are reported, if not embedded in the privacy policy directly                                                                                                                                                                                                                                                                                                                                                                             &  \\
Date collected                                                                                                      & MM/DD/YYYY              & Date the privacy policy was reviewed for metric reporting                                                                                                                                                                                                                                                                                                                                                                                                                                   &  \\
Date policy last updated                                                                                            & MM/DD/YYYY or none      & Reported date of last update on broker privacy policy, none if not reported                                                                                                                                                                                                                                                                                                                                                                                                                 &  \\
\begin{tabular}[c]{@{}l@{}}Total \# Requests to Know\\ Collecting\end{tabular}                                      & integer or none         & \begin{tabular}[c]{@{}l@{}}Reported number of requests from consumers in 2024 to know what data the data broker is collecting on them  (none if none reported,\\ none if the requests are not explicitly from 2024). If the “right to know what broker is selling/sharing” is included \\ in the description of the “right to know” in the policy, report same metrics for both requests to know what is being collected \\ and requests to know what is being sold/shared.\end{tabular}    &  \\
\begin{tabular}[c]{@{}l@{}}Complied with in whole or \\ in part \# Requests to Know\\ Collecting\end{tabular}       & integer or none         & Reported number of complied requests from consumers in 2024 on the request to know what the data broker is collecting on them                                                                                                                                                                                                                                                                                                                                                               &  \\
\begin{tabular}[c]{@{}l@{}}Denied \# Requests to Know\\ Collecting\end{tabular}                                     & integer or none         & Reported number of denied requests from consumers in 2024 on the request to know what the data broker is collecting on them                                                                                                                                                                                                                                                                                                                                                                 &  \\
Requests to know combined?                                                                                          & TRUE / FALSE            & \begin{tabular}[c]{@{}l@{}}If the data broker explicitly states in their definition of ``right to know'' that it combines both the right to know what data is \\ being collected, as well as being sold and shared mark TRUE. Else, mark FALSE.\end{tabular}                                                                                                                                                                                                                                  &  \\
\begin{tabular}[c]{@{}l@{}}Total \# Requests to Know\\ Selling/ sharing\end{tabular}                                & integer or none         & \begin{tabular}[c]{@{}l@{}}Reported number of requests from consumers in 2024 to know what data the data broker is selling/sharing about them  (none if none reported,\\ none if the requests are not explicitly from 2024). If the “right to know what broker is selling/sharing” is included \\ in the description of the “right to know” in the policy or if Requests to know combined? = TRUE, this should be the same as \\ Total \# Requests to Know Collecting integer)\end{tabular} &  \\
\begin{tabular}[c]{@{}l@{}}Complied with in whole or \\ in part \# Requests to Know\\ Selling/ sharing\end{tabular} & integer or none         & Reported number of complied requests from consumers in 2024 on the request to know what the data broker is selling / sharing about them                                                                                                                                                                                                                                                                                                                                                     &  \\
\begin{tabular}[c]{@{}l@{}}Denied \# Requests to Know\\ Selling / sharing\end{tabular}                              & integer or none         & Reported number of denied requests from consumers in 2024 on the request to know what the data broker is selling / sharing about them                                                                                                                                                                                                                                                                                                                                                       &  \\
Total \# Requests to Delete                                                                                         & integer or none         & Reported number of requests from consumers in 2024 to delete their data from the broker                                                                                                                                                                                                                                                                                                                                                                                                     &  \\
\begin{tabular}[c]{@{}l@{}}Complied with in whole or \\ in part \# Requests to Delete\end{tabular}                  & integer or none         & Reported number of complied requests from consumers in 2024 on the request to delete their data from the broker                                                                                                                                                                                                                                                                                                                                                                             &  \\
Denied \# Requests to Delete                                                                                        & integer or none         & Reported number of denied requests from consumers in 2024 on the request to delete their data from the broker                                                                                                                                                                                                                                                                                                                                                                               &  \\
\begin{tabular}[c]{@{}l@{}}Total \# Requests to Do Note Sell\\ (Opt Out)\end{tabular}                               & integer or none         & Reported number of requests from consumers in 2024 for the data broker to not sell their data (or, opt out of data collection / sharing)                                                                                                                                                                                                                                                                                                                                                    &  \\
\begin{tabular}[c]{@{}l@{}}Complied with in whole or \\ in part \# Requests to Do Not Sell\\ (Opt Out)\end{tabular} & integer or none         & \begin{tabular}[c]{@{}l@{}}Reported number of complied requests from consumers in 2024 on the request that data broker \\ not sell their data (or, opt out of data collection / sharing)\end{tabular}                                                                                                                                                                                                                                                                                       &  \\
\begin{tabular}[c]{@{}l@{}}Denied \# Requests to Do Not Sell\\ (Opt Out)\end{tabular}                               & integer or none         & \begin{tabular}[c]{@{}l@{}}Reported number of denied requests from consumers in 2024 on the request that data broker \\ not sell their data (or, opt out of data collection / sharing)\end{tabular}                                                                                                                                                                                                                                                                                         &  \\
Total \# Requests to Limit                                                                                          & integer or none         & Reported number of requests from consumers in 2024 for the data broker to limit the sharing / selling of their sensitive personal information                                                                                                                                                                                                                                                                                                                                               &  \\
\begin{tabular}[c]{@{}l@{}}Complied with in whole or \\ in part \# Requests to Limit\end{tabular}                   & integer or none         & \begin{tabular}[c]{@{}l@{}}Reported number of complied requests from consumers in 2024 on the request that  the data broker limit the \\ sharing / selling of their sensitive personal information\end{tabular}                                                                                                                                                                                                                                                                             &  \\
Denied \# Requests to Limit                                                                                         & integer or none         & \begin{tabular}[c]{@{}l@{}}Reported number of denied requests from consumers in 2024 on the request that  the data broker limit the \\ sharing / selling of their sensitive personal information\end{tabular}                                                                                                                                                                                                                                                                               &  \\
Explicit location mention                                                                                           & CA, US, GLOBAL, or none & \begin{tabular}[c]{@{}l@{}}Report CA if the broker explicitly states the metrics are from Californian consumers. Report US if they state that they are from US consumers.\\ Report GLOBAL if they explicitly state they are from any location. Report none if no explicit mention of the location scope of \\ consumer request metrics.\end{tabular}                                                                                                                                        &  \\
Mean/median reported?                                                                                              & TRUE / FALSE            & Report TRUE if the median or mean time to fulfill requests is reported, else FALSE.                                                                                                                                                                                                                                                                                                                                                                                                         &  \\
All requests reported?                                                                                              & TRUE / FALSE            & After reviewing for all requests, were all five request metrics reported?                                                                                                                                                                                                                                                                                                                                                                                                                   &  \\ \bottomrule
\end{tabular}%
}
\end{table}

\FloatBarrier
\label{sec:codebook}
\subsection{Silent Duplicates}
\label{sec:silentdup}
Following Gerchick et al. \cite{gerchick-2025} who define silent duplicates in the context of New York's Local Law 144, we use the term ``silent duplicates'' to refer to independently registered data brokers with identical or nearly identical privacy policies.  Here we offer, albeit anecdotal, evidence of a nontrivial number of silent duplicates in the 2025 California Data Broker Registry, many of which obscure their corporate relationships with one another. As documented in Table \ref{tab:silentdups}, we identify several instances in which brokers report identical 2024 request metrics, share identical contact information, or use significantly similar language in their policies, yet do not reference one another as affiliates, subsidiaries, or DBAs in either the registry nor their public privacy policies. This lack of transparency hinders the consumer's ability to understand which broker holds their data and whether a given request will be honored across related entities. 

We observe that silent duplicates are particularly common among people search websites (PSWs), a category of data brokers that have been previously shown to be opaquely connected and largely noncompliant with consumer access requests \cite{take-2024}. Though they are registered as separate entities, multiple PSWs have identical privacy policies and website design, but do not reference each other. We acknowledge that there are several benign explanations for silent duplicates: brokers may have the same lawyer, use the same privacy policy template as one another as it is easier than creating their own, or operate under a recent corporate merging or acquisition that has not yet been documented in the registry. Moreover, we note that many silent duplicates appear close to one another in the registry's ordering, which may suggest coordinated registration by the same individual or entity, though we do not know for sure as the CPPA does not explicitly say how they determine the registry order. However, there may also be more opaque reasons for duplicates, such as corporate structure benefits from having multiple related brokers like increased data sharing or the diffusion of opt-out or deletion requests across nominally distinct brokers.

Finally, silent duplicates may increase consumer burden even when brokers provide identical contact information. In such cases, it is not easily discernible whether a given request via a shared email address will apply to all related brokers or to a single entity and, if so, which broker it would apply to. Some brokers explicitly state in their privacy policies that certain consumer requests will be applied across all affiliates or subsidiaries, while others require consumers to submit requests separately to each registered broker. This inconsistency in honoring consumer requests further infringes on consumer privacy rights. The abundance of silent duplicates demonstrates the need for clearer disclosure of corporate relationships both on the registry and within privacy policies.

Though our documentation of silent duplicates was purely observational, one could potentially automate the identification of duplicates by quantitatively measuring the similarity of privacy policies using NLP methods (such as cosine similarity of two document TF-IDF  vectors or greedy longest common subsequence), then reviewing policies with high similarity scores. The data broker registry moreover requires brokers to provide the link to their privacy policy, rights request metrics, contact information, and primary address. Theoretically, this would enable researchers to quickly identify duplicate information within the registry. However, in practice, we found the registry to be inconsistent with the data we manually collected from the brokers' websites, and contained several broken links.

 Although the recent implementation of DROP may mitigate some of these concerns by allowing consumers to submit deletion requests to all registered brokers simultaneously, it remains unclear how requests are communicated and honored across duplicates. Given how difficult it is currently to ascertain why exactly these duplicates exist, we hope future research will further investigate this issue and how the prevalence of silent duplicates may change now that DROP has gone into effect.

%Opaqueness in providing the same privacy policy link, all entities report the same metrics, so consumers can't tell which request goes to which entity or who holds what, though this is a lesser concern now that DROP allows you to delete from all brokers on the registry.

%Consumer burden possibly increased because of entities same contact information, don't know who request is being submitted to or if data is being deleted from all entities. Some brokers specifically say in their privacy policy, such as Mississippi Tornado Alley LLC, that the some types of consumer requests will be applied across all affiliates/subsidiaries, while others require consumers to opt out on each individual website.

%"Mississippi Tornado Alley LLC operates the following websites: Usapeoplesearch.com; Advancedbackgroundchecks.com; Peoplesearchnow.com; USphonebook.com; Fastpeoplesearch.com; Searchpeoplefree.com; Smartbackgroundchecks.com; Cyberbackgroundchecks.com; Fastbackgroundcheck.com. To learn about the privacy practices for each site, please check the privacy policy located in the footer of each website.  Consumers can exercise their right to opt out of the sale of certain personal information by submitting a request on any one of these sites. The opt out request will be applied across all of these sites. To opt out of the sale or sharing of personal information that occurs through the online collection technologies we use (e.g., cookies), consumers will need to opt out separately on each website."

\setlength{\extrarowheight}{2pt}
\begin{table}[t]
\caption{\textbf{Examples of silent duplicates among data brokers in the 2025 California registry.} This table documents independently registered brokers with identical or nearly identical privacy policies. Broker names are written exactly as listed in the registry. ``Ref. each other'' indicates whether at least one broker in the group references another in its privacy policy or is listed as DBA in the registry. ``Same metrics'' indicates whether brokers report identical request numbers. ``Same policy link'' indicates if the brokers provide the same link to their respective privacy policies. ``Identical contact info'' lists which, if any, forms of contact information (phone, mailing address, email address) are identical, as listed in the registry or privacy policies. ``Same state'' documents whether the duplicates are located in the same U.S. state and, if so, which state that is. Finally, ``Notes'' summarizes any additional qualitative observations from the manual review. Here, ``DBA'' stands for ``Doing business as'' and ``PSW'' refers to people search website.}
\label{tab:silentdups}
\centering
\SMALL
\renewcommand{\arraystretch}{1.08} % a bit more row height (optional)
\resizebox{\columnwidth}{!}{%
\begin{tabular}{p{0.28\columnwidth} c c c p{0.15\columnwidth} c p{0.27\columnwidth}}
\toprule
\textbf{List of Silent Duplicates} &
\textbf{Ref. each other} &
\textbf{Same metrics} &
\textbf{Same policy link} &
\textbf{Identical contact info} &
\textbf{Same State} &
%\textbf{Similar website} &
\textbf{Notes} \\
\midrule
Sabio Inc.; AppScience                                                                                                                                                                              & No                   & No           & No                       & Phone, Mailing address                 & Yes (CA)                       & Mentions each other on respective websites, but not in privacy policy nor registry.                             \\\hline

Matchbook Data, LLC; Outlogic, LLC                                                                                                                                                                  & No                   & Yes          & No                       & Mailing address                        & Yes (NY)              &    \\\hline

Unacast, Inc.; Venntel, Inc.                                                                                                                                                                  & No                   & N/A          & No                       & None                        & Yes (VA)             & No metrics reported for either broker.                           \\\hline
Intelius, LLC; TruthFinder, LLC; Instant Checkmate, LLC                                                                                                                                                            & No                   & No           &  No                       &     None                                   & Yes (CA)                 & All are PSWs with nearly identical websites located in San Diego.                               \\\hline
J2 Global Canada, Inc.; J2 Martech Corp.                                                                                                                                                            & No                      &     Yes         & Yes                      &  None                                      &    No                   & One located in Canada, other located in Delaware, USA. Only J2 Martech Corp. registered DBA: ``Full Contact''.                              \\\hline
eXelate Inc.; The Nielsen Company                                                                                                                                                                   &  Yes                    &     Yes         &     Yes                     &    Phone, Mailing address, Email                                   & Yes (NY)                   &  Both registered DBA: ``Nielsen Marketing Cloud.''
\\\hline
We Inform LLC; Truth Now LLC; Private Records LLC; The People Searchers LLC; Infomatics LLC                                                                                                          & No                   & No           & No                       & None                                   & No                    & All are PSWs with nearly identical websites. Two located in CA, three located in FL. \\\hline
Consumerbase, LLC; DonorBase, Inc.; Data Axle Inc.                                                                                                                                                    &      Yes                &   Yes           &  Yes                        &  Phone, Mailing address                                      & Yes (TX)                 &  All listed as affiliate or subsidiary in privacy policy.        \\\hline
PeopleFinders LLC; Mississippi Tornado Alley LLC; Free Data Services, LLC; Family Tree Now; LLC                                                                    &   No                   & No             &        No                  &   None                                     &   Yes (CA)        &  All are PSWs with nearly identical websites. Last three brokers in list all have same mailing address.                             \\\hline
Digital Safety Products, LLC.; National Data Analytics, LLC; Information Data Resources, LLC; Civil Data Research, LLC; Scalable Commerce, LLC                                                      &     No                 &    No          &     No                     &    No                                    &    Yes (CA)                        & All are PSWs with nearly identical websites. Registered other companies as DBA, but did not reference each other.                              \\\hline
CheckPeople, LLC; Unmask, LLC; FreePeopleSearch.com, LLC                                                                                                                                                           &    No                  &   No           &  No                        & No                                       &    Yes (FL)                        & All are PSWs with nearly identical websites.                       \\\hline
33 Mile Radius LLC; Keyword Connects LLC; Remodelling.com, LLC; Best Pick Reports, LLC; Home Contractors Review, LLC                                                                                                                                                        &    No                  &   Yes          &  No                        &  No                                      &  No                        & %First four brokers registered DBA: ``EverConnect'' but Home Contractors Review, LLC did not. 
First three located in CO, last two located in GA. Do not reference each other except for Remoddeling.com, LLC's privacy policy references itself, 33 Mile Radius, LLC and Keyword Connects LLC as DBA EverConnect.                         \\\hline
Complete Medical Lists, Inc.; Complete Mailing Lists LLC                                                                                       &  Yes                    &  No            &  No                        &    No                                    &  No                        &  Former located in NH, latter in NY.                              \\\hline
TransUnion LLC; TransUnion Content Solutions LLC; TransUnion Digital LLC; TransUnion Risk and Alternative Data Solutions, Inc.; Tru Optik Data Corp.; TruSignal, Inc.; TransUnion Interactive, Inc. &  Yes                    &   Yes           &                Yes          & Phone, Mailing address, Email                                             &   Yes (IL)              & Do not reference each other in registry, but some listed as affiliates in privacy policy.                              \\\hline
Intalytics, Inc.; eSite Analytics Inc                                                                                    &  Yes                   &  N/A            & Yes                      &   Email                                        &  No               &  Both registered DBA: ``Kalibrate''. Former located in MI, latter in SC. No metrics reported.                           \\\hline
Swordfish AI Inc.; Heartbeat.AI Inc   &  No                    &   Yes           &  No                        &                         Phone, Mailing address             &    Yes (DE)                  &  Heartbeat.ai website says ``Powered by Swordfish.ai'' but brokers do not reference each other in privacy policies nor registry.                             \\\hline
Tunnl, LLC; Deep Root Analytics, LLC                                                                                                                                  & No                     & Yes             & No                         & Phone, Mailing address, Email                                       &  Yes (VA)                        &  Do not reference each other, but Deep Root Analytics lists same contact email as Tunnl in their privacy policy.                            \\\hline
Compact Information Systems, LLC (4 times)                                                                                                                                                                &   Yes                  &  N/A            &  Yes                        &   Phone, Mailing address, Email                                    &   Yes (WA)                 &  Broker was listed in the registry four times with different DBA entries. No reported metrics. DBA names listed as affiliates in policy.                 \\\hline
Austin Consolidated Holdings, Inc.; Compliance Data Center LLC; IXI Corporation; Knowledge Works, Inc.; Equifax Workforce Solutions LLC; Equifax Information Services LLC                           &  No                    &  Yes            &  Yes                        &  Phone, Mailing address, Email                                                 & Yes (GA)                &  Same link, but do not list each other as DBA in the registry nor as affiliates/subsidiaries in the privacy policy.                            \\\hline
Experian Marketing Solutions, LLC; Experian Information Solutions, Inc.; Experian Health, Inc.                                                                                                      &  Yes                    &   Yes           &  Yes                        &  Phone, Mailing address, Email                                      &  No                      &  All three located in different states: IL, CA, and TN, respectively. All referenced as subsidiaries in privacy policy.                        
\\
\bottomrule 
\end{tabular}
}
\end{table}

\FloatBarrier

\subsection{Description of Predictor Variables}
\label{sec:predictor-variables}
In order to uncover more information about the variables that may correlate with which data brokers report metrics as well as the actual amount of requests they receive, we conducted a descriptive analysis between categorical data based on the 2025 Data Broker Registry and via data obtained from Dun \& Bradstreet via Stanford University's library services. All variables investigated as correlates to our questions of interests are listed with data type and description in Table \ref{tab:predictor-variables}.

\begin{table}[h]
\caption{Codebook of predictor variables used in modeling requests received / if requests are reported}
\label{tab:predictor-variables}
\resizebox{\columnwidth}{!}{%
\begin{tabular}{@{}lll@{}}
\toprule
Predictor variable                                                           & Data type & Description                                                                                                                         \\ \midrule
Income                                                                       & Integer   & Annual income from 2022 (most recent data we could obtain)\\
Employees                                                                    & Integer   & Number of employees across all branches                                                                                             \\
CA located                                                                   & 0 / 1     & Whether the data broker is based in California (self-reported state location from the data broker)                                  \\
\begin{tabular}[c]{@{}l@{}}Collects data from\\ minors\end{tabular}          & 0 / 1     & Does the data broker collect information from minors?                                                                               \\
\begin{tabular}[c]{@{}l@{}}Collects precise \\ geolocation data\end{tabular} & 0 / 1     & Does the data broker collect consumers’ precise geographic locations?                                                               \\
\begin{tabular}[c]{@{}l@{}}Collects reproductive\\ health data\end{tabular}  & 0 / 1     & Does the data broker collect consumers’ reproductive information?                                                                   \\
Subject to CMIA                                                              & 0 / 1     & The data broker or any of its subsidiaries is regulated by the California Confidentiality of Medical Information Act (CMIA)?        \\
Subject to IIPPA                                                             & 0 / 1     & The data broker or any of its subsidiaries is regulated by the California Insurance Information and Privacy Protection Act (IIPPA)? \\
Subject to HIPAA                                                             & 0 / 1     & The data broker or any of its subsidiaries is regulated by the Health Insurance Portability and Accountability Act (HIPAA)?         \\
Subject to FCRA                                                              & 0 / 1     & The data broker or any of its subsidiaries is regulated by the federal Fair Credit Reporting Act (FCRA)?                            \\
Subject to GLBA                                                              & 0 / 1     & The data broker or any of its subsidiaries is regulated by the federal Gramm-Leach-Bliley Act (GLBA)?                               \\
Is subsidiary                                                                & 1 / 3     & The data broker is a subsidiary (3) or a parent (1) (the 1 and 3 are defined by Dun \& Bradstreet data)                             \\ \bottomrule
\end{tabular}%
}
\end{table}
\FloatBarrier

\clearpage
\subsection{Timeline of Data Broker Registration and Reporting}
\label{reporting-timeline}
Data brokers must register annually with the California Privacy Protection Agency (CPPA) by January 31 and submit consumer request metrics from two calendar years prior. Brokers must also publicly post rights requests metrics from the previous calendar year by July 1 to their online privacy policies. Table ~\ref{tab:reporting_timeline} summarizes these reporting deadlines and the relevant regulatory developments.
\begin{table}[h]
\centering
\caption{Timeline of data broker reporting and enforcement requirements under the CCPA and Delete Act.}
\label{tab:reporting_timeline}
\begin{tabular}{p{0.5cm}p{4.5cm}p{4.5cm}p{4.25cm}}
\toprule
Year & Data Broker Registry Reporting & Transparency Requirement & Regulatory Developments \\
\midrule

2025 & \textbf{Jan 31:} Brokers register and submit \textbf{2023 request metrics} to CPPA & \textbf{July 1:} Brokers publicly post \textbf{2024 request metrics} on websites & Transparency reporting requirement takes effect. \\
2026 & \textbf{Jan 31:} Brokers register and submit \textbf{2024 request metrics} to CPPA & \textbf{July 1:} Brokers publicly post \textbf{2025 request metrics} on websites & DROP platform operational (January 1st, 2026). \\
2027 & \textbf{Jan 31:} Brokers register and submit \textbf{2025 request metrics} to the CPPA& \textbf{July 1:} Brokers must publicly post \textbf{2026 request metrics} on websites & California Opt Me Out Act goes into effect (January 1st, 2027). \\

2028 & \textbf{Jan 31:} Brokers register and submit \textbf{2025 request metrics} to the CPPA & \textbf{July 1:} Brokers must publicly post \textbf{2026 request metrics} on websites & First compliance audit cycle begins (January 1st, 2028).\\

\bottomrule
\end{tabular}
\end{table}

\clearpage
\subsection{Additional Analyses}
\label{logistic-regression-appendix}
Table \ref{tab:coefficients-from-regressions} shows results from the logistic regression coefficients for data brokers that report metrics versus those that don't report. The remaining tables (11 through 14) investigate the difference between variables of data brokers and the number of specific requests they report. 

\begin{table}[H]
\caption{Table of coefficients from logistic regression. P-values calculated using ANOVA. Only being a corporate subsidiary is significantly associated with whether or not a data broker reports.}
\label{tab:coefficients-from-regressions}
%\resizebox{0.8\textwidth}{!}
\scalebox{0.9} {%
\begin{tabular}{@{}lll@{}}
\toprule
                                                                              & Coeff. estimate & p-value \\ \midrule
Income                                                                        & 0.05            & 0.819 \\
Employees                                                                     & 0.18            & 0.674 \\
CA located                                                                    & 2.35            & 0.125 \\
\begin{tabular}[c]{@{}l@{}}Collects data\\ from minors\end{tabular}           & 1.17            & 0.280 \\
\begin{tabular}[c]{@{}l@{}}Collects precise\\ geolocation\\ data\end{tabular} & 0.18            & 0.673 \\
\begin{tabular}[c]{@{}l@{}}Collects reproductive\\ health data\end{tabular}   & 0.55            & 0.456 \\
Subject to CMIA                                                               & 0.14            & 0.704 \\
Subject to IIPPA                                                              & 0.88            & 0.346 \\
Subject to HIPAA                                                              & 0.03            & 0.875 \\
Subject to FCRA                                                               & 3.00            & 0.084 \\
Subject to GLBA                                                               & 0.64            & 0.425 \\
Is subsidiary                                                                 & \textbf{4.00 *} & 0.046 \\ \bottomrule
\end{tabular}%
}
\end{table}
%% remaining tables of differences for each individual right 

% \usepackage{booktabs}
% \usepackage{graphicx}
\begin{table}[H]
\caption{Comparison of correlates between data brokers receiving low and high numbers of requests to delete.  *Denotes a subgroup of low  (N=65) and high reporting (N=66) brokers due to unavailable corporate entity data. **P-values for the proportion of data brokers with a certain characteristic is calculated using a two sample proportion test while p-values for corporate income and number employees are calculated using t-tests.}
\label{tab:right-to-delete-differences}
\resizebox{\columnwidth}{!}{%
\begin{tabular}{@{}llllll@{}}
\toprule
                                        & \begin{tabular}[c]{@{}l@{}}Low requests received \\ (N=139)\end{tabular} &              & \begin{tabular}[c]{@{}l@{}}High requests received\\ (N=139)\end{tabular} &              &               \\ \midrule
                                        & Percentage                                                               & SE           & Percentage                                                               & SE           & p-value**       \\ \cmidrule(l){2-6} 
Collects data from minors (\%)          & 2.2                                                                      & 1.2          & 5.6                                                                      & 2.0          & 0.22          \\
Collects precise geolocation data  (\%) & 19.4                                                                     & 3.3          & 12.9                                                                     & 2.8          & 0.13          \\
Collects reproductive health data (\%)  & 1.4                                                                      & 1.1          & 1.4                                                                      & 1.1          & 1             \\
Subject to FCRA (\%)                    & 1.4                                                                      & 1.0          & 4.2                                                                      & 1.7          & 0.28          \\
Subject to GLBA (\%)                    & 4.3                                                                      & 1.7          & 4.3                                                                      & 1.7          & 1             \\
Subject to CMIA (\%)                    & 1.4                                                                      & 1.0          & 0                                                                        & 0            & 0.47          \\
Subject to IIPPA (\%)                   & 0.72                                                                     & 0.72         & 0                                                                        & 0            & 0.99          \\
Subject to HIPAA (\%)                   & 4.3                                                                      & 1.7          & 6.4                                                                      & 2.1          & 0.60          \\
From California (\%)                    & 23.0                                                                     & 3.5          & 21.6                                                                     & 3.5          & 0.80          \\
Is subsidiary*                & 30.2                                                          & 5.8 & 28.1                                                            & 5.6 & 0.95 \\ \midrule
                                        & Mean                                                                     & SE           & Mean                                                                     & SE           & p-value**       \\ \cmidrule(l){2-6} 
Income*                                 & 341,922,387                                                             & 218,601,422  & 469,220,118                                                              & 427,190,755  & 0.79          \\
Employees*                              & 2,070                                                                    & 2,551        & 890                                                                   & 548          & 0.41          \\ \bottomrule
\end{tabular}%
}
\end{table}

% \usepackage{booktabs}
% \usepackage{graphicx}
\begin{table}[H]
\caption{Comparison of correlates between data brokers receiving low and high numbers of "do not sell requests". *Denotes a subgroup of data brokers with lower (N=63) and higher (N=63) request amounts is due to unavailable corporate entity data for all brokers  reporting "do not sell" requests. **P-values for the proportion of data brokers with a certain characteristic is calculated using a two sample proportion test while p-values for corporate income and number of employees are calculated using t-tests.}
\label{tab:do-not-sell-differences}
\resizebox{\columnwidth}{!}{%
\begin{tabular}{@{}llllll@{}}
\toprule
                                        & \begin{tabular}[c]{@{}l@{}}Low requests received \\ (N=139)\end{tabular} &            & \begin{tabular}[c]{@{}l@{}}High requests received\\ (N=140)\end{tabular} &             &         \\ \midrule
                                        & Percentage                                                               & SE         & Percentage                                                               & SE          & p-value** \\ \cmidrule(l){2-6} 
Collects data from minors (\%)          & 4.1                                                                      & 1.7        & 3.6                                                                      & 1.6         & 0.99    \\
Collects precise geolocation data  (\%) & 15.1                                                                     & 3.0        & 17.1                                                                     & 3.2         & 0.76    \\
Collects reproductive health data (\%)  & 0.72                                                                     & 0.72       & 2.1                                                                      & 1.2         & 0.62    \\
Subject to FCRA (\%)                    & 0.72                                                                     & 0.72       & 5.0                                                                      & 1.8         & 0.07    \\
Subject to GLBA (\%)                    & 3.6                                                                      & 1.6        & 5.7                                                                      & 2.0         & 0.58    \\
Subject to CMIA (\%)                    & 0.72                                                                     & 0.72       & 0.71                                                                     & 0.71        & 1       \\
Subject to IIPPA (\%)                   & 0.72                                                                     & 0.72       & 0                                                                        & 0           & 0.99    \\
Subject to HIPAA (\%)                   & 5.0                                                                      & 1.9        & 5.7                                                                      & 2.0         & 1       \\
From California (\%)                    & 22.3                                                                     & 3.5        & 21.0                                                                     & 3.4         & 0.86    \\
Is subsidiary*                          & 27.0                                                                     & 4.9        & 31.2                                                                     & 5.8         & 0.70    \\ \midrule
                                        & Mean                                                                     & SE         & Mean                                                                     & SE          & p-value** \\ \cmidrule(l){2-6} 
Income*                                 & 68,661,019                                                               & 29,174,003 & 749,900,723                                                              & 482,885,254 & 0.16    \\
Employees*                              & 492                                                                      & 264        & 2,964                                                                    & 1,969       & 0.21    \\ \bottomrule
\end{tabular}%
}
\end{table}

% \usepackage{booktabs}
% \usepackage{graphicx}
\begin{table}[H]
\caption{Comparison of correlates between data brokers reporting zero and non-zero "requests to limit sharing of sensitive personal information".  *Denotes a subgroup of data brokers reporting zero (N=61) and a non-zero (N=26)  number of requests due to unavailable corporate entity data for all brokers reporting requests to limit. **P-values for proportion of data brokers that have a certain characteristic is calculated using a two sample proportion test while p-values for corporate income and number of employees are calculated using t-tests.}
\label{tab:right-to-limit-differences}
\resizebox{\columnwidth}{!}{%
\begin{tabular}{@{}llllll@{}}
\toprule
                                        & \begin{tabular}[c]{@{}l@{}}Zero requests received\\ (N = 126)\end{tabular} &             & \begin{tabular}[c]{@{}l@{}}More than zero requests\\ received (N = 61)\end{tabular} &              &                 \\ \midrule
                                        & Percentage                                                                 & SE          & Percentage                                                                          & SE           & p-value         \\ \cmidrule(l){2-6} 
Collects data from minors (\%)          & 4.8                                                                        & 1.9         & 8.2                                                                                 & 3.5          & 0.55            \\
Collects precise geolocation data  (\%) & 16.7                                                                       & 3.3         & 11.5                                                                                & 4.1          & 0.48            \\
Collects reproductive health data (\%)  & 0                                                                          & 0           & 3.3                                                                                 & 2.3          & 0.20            \\
\textbf{Subject to FCRA (\%)}           & \textbf{0}                                                                 & \textbf{0}  & \textbf{11.5}                                                                       & \textbf{4.1} & \textbf{0.0005} \\
Subject to GLBA (\%)                    & 4.0                                                                        & 1.7         & 11.5                                                                                & 4.1          & 0.10            \\
Subject to CMIA (\%)                    & 1.6                                                                        & 1.1         & 0                                                                                   & 0            & 0.82            \\
Subject to IIPPA (\%)                   & 0.79                                                                       & 0.79        & 0                                                                                   & 0            & 1               \\
Subject to HIPAA (\%)                   & 4.7                                                                        & 1.9         & 8.2                                                                                 & 3.5          & 0.55            \\
From California (\%)                    & 19.8                                                                       & 3.6         & 21.3                                                                                & 5.2          & 0.97            \\
Is subsidiary*                          & 30.0                                                                       & 5.8         & 20.6                                                                                & 7.9          & 0.1538          \\ \midrule
                                        & Mean                                                                       & SE          & Mean                                                                                & SE           & p-value         \\ \cmidrule(l){2-6} 
Income*                                 & 765,717,623                                                                & 499,125,219 & 114,114,359                                                                         & 37,700,591   & 0.20            \\
Employees*                              & 3,088                                                                      & 2,046       & 766                                                                                 & 300          & 0.27            \\ \bottomrule
\end{tabular}%
}
\end{table}

% \usepackage{booktabs}
% \usepackage{graphicx}

\begin{table}[H]
\caption{Comparison of correlates between data brokers receiving low and high numbers of  "requests to know what is being collected" and "requests to know what is being sold".  *Denotes a subgroup of data brokers with higher (N=61) and lower (N=63) request amounts due to unavailable corporate entity data for all brokers reporting requests to limit. **P-values for proportion of data brokers with a certain characteristic is calculated using a two sample proportion test while p-values for corporate income and number of employees are calculated using t-tests.}
\label{tab:right-to-know-differences}
\resizebox{\columnwidth}{!}{%
\begin{tabular}{@{}llllll@{}}
\toprule
                                        & \begin{tabular}[c]{@{}l@{}}Low requests received\\ (N = 138)\end{tabular} &              & \begin{tabular}[c]{@{}l@{}}High requests received\\  (N = 136)\end{tabular} &              &               \\ \midrule
                                        & Percentage                                                                & SE           & Percentage                                                                  & SE           & p-value**     \\ \cmidrule(l){2-6} 
\textbf{Collects data from minors (\%)} & \textbf{0.7}                                                              & \textbf{0.7} & \textbf{7.4}                                                                & \textbf{2.2} & \textbf{0.01} \\
Collects precise geolocation data  (\%) & 18.8                                                                      & 3.3          & 13.2                                                                        & 2.9          & 0.27          \\
Collects reproductive health data (\%)  & 2.2                                                                       & 1.2          & 0.7                                                                         & 0.7          & 0.62          \\
\textbf{Subject to FCRA (\%)}           & \textbf{0}                                                                & \textbf{0}   & \textbf{5.9}                                                                & \textbf{2.0} & \textbf{0.01} \\
\textbf{Subject to GLBA (\%)}           & \textbf{1.4}                                                              & \textbf{1.0} & \textbf{7.4}                                                                & \textbf{2.2} & \textbf{0.03} \\
Subject to CMIA (\%)                    & 0                                                                         & 0            & 1.5                                                                         & 1.0          & 0.47          \\
Subject to IIPPA (\%)                   & 0                                                                         & 0            & 0.7                                                                         & 0.7          & 0.99          \\
Subject to HIPAA (\%)                   & 3.6                                                                       & 1.6          & 7.3                                                                         & 2.2          & 0.27          \\
From California (\%)                    & 21.0                                                                      & 3.5          & 22.3                                                                        & 3.6          & 0.83          \\
\textbf{Is subsidiary*}                 & \textbf{19.0}                                                             & \textbf{4.9} & \textbf{41.0}                                                               & \textbf{6.3} & \textbf{0.02} \\ \midrule
                                        & Mean                                                                      & SE           & Mean                                                                        & SE           & p-value**     \\ \cmidrule(l){2-6} 
Income*                                 & 73,124,481                                                                & 31,730,270   & 769,903,595                                                                 & 498,472,043  & 0.17          \\
Employees*                              & 505                                                                       & 267          & 3,048                                                                       & 2,033        & 0.22          \\ \bottomrule
\end{tabular}%
}
\end{table}

\subsection{Comparison to 2026 California Data Broker Registry}
\label{2026-registry-appendix}
As an additional robustness check, we compared our hand-collected dataset of 2025 website-posted privacy request metrics to the centralized metrics reported in the 2026 California Data Broker Registry, posted in April 2026. This comparison allows us to assess whether centralized reporting improves the availability of request metrics and whether the numbers reported across sources are internally consistent. We matched brokers by exact name, normalized name, DBA, and known aliases, yielding 457 matched brokers with the 2025 dataset. Because the two sources may differ in reporting timing, formatting, and interpretation of missing values, we interpret discrepancies as evidence of reporting inconsistency rather than definitive proof that either source is incorrect. That said, per statutory requirements the metrics posted to brokers' websites as of July 1, 2025 should be identical to those posted to the registry as of Jan. 2026, given that they span the exact same time period (Jan.-Dec. 2024). 

The comparison supports our finding that decentralized website reporting makes systematic monitoring difficult. Only 45 brokers (9.8\%) reported the same metric coverage across both sources and only 27 (5.9\%) brokers had identical reported metrics across both sources (Table ~\ref{tab:registry-comparison}). Overall, the 2026 registry reported more of the core request-count fields (for the five rights request categories) than our hand-collected dataset for 412 matched brokers (90.2\%), suggesting that brokers under-reported their metrics reporting on their privacy policies. For 199 matched brokers (43.5\%), our hand-coded dataset contained no request-count metrics while the 2026 registry reported at least one metric, which suggests that some brokers did not take the July 2025 deadline seriously and failed to report. There were 315 matched brokers (68.9\%) with ``missing-only differences'', which we use describe brokers whose discrepancies consisted entirely of the 2026 registry reporting a request-count metric where our hand-coded dataset was coded as none/missing (in part because missing values in website-reporting are generally replaced with zeros when registering with the CPPA); for these brokers, there were no fields where both sources reported different numeric values. In other words, these brokers differed in whether metrics were reported at all, not in the value of any metric reported by both sources. Because populated fields in the registry include those with zero reports we also examined nonzero request-count fields: excluding fields with zeros, the 2026 registry contained more nonzero request-count fields than our hand-collected dataset for 320 matched brokers (70.0\%). In all, these results suggest that centralized registry reporting substantially improves metric availability.

There were also issues with numeric consistency between sources. Among matched brokers, 115 brokers (25.2\%) had at least one true numeric mismatch where both the hand-collected source and the 2026 registry reported a value, including zero. Across all 494 numeric mismatched cells, 271 differences (54.9\%) reflected higher values in the 2026 registry than in the hand-coded data, while 223 differences (45.1\%) reflected lower values in the registry (Table~\ref{tab:registry-direction}). Thus, by count, the registry value was more often higher than the website-posted value, indicating that website-posted metrics generally under-counted when compared to the corresponding registry values. However, the net signed difference was positive because several large opt-out discrepancies made the hand-coded totals larger. The median mismatch sizes were much smaller than the means, indicating that the average discrepancy is strongly influenced by large outliers. But there were 77 brokers who reported at least one lower metric to the registry than they posted on their own websites, with 52 (11\%) of those specifically reporting lower total request values. The most significant cases of underreporting requests were Tapad (over 29 million reports posted as compared to 2,537 in the registry), T-Mobile USA (over 16 million reports posted as compared to 1.6 million in the registry), and AtData (1.4 million reports posted as compared to 136K in the registry). We have no explanation for why some brokers reported lower numbers to the registry than they reported on their own websites, given that the coverage period should have been identical. 

\begin{figure}[t]
    \centering
    \includegraphics[width=\linewidth]{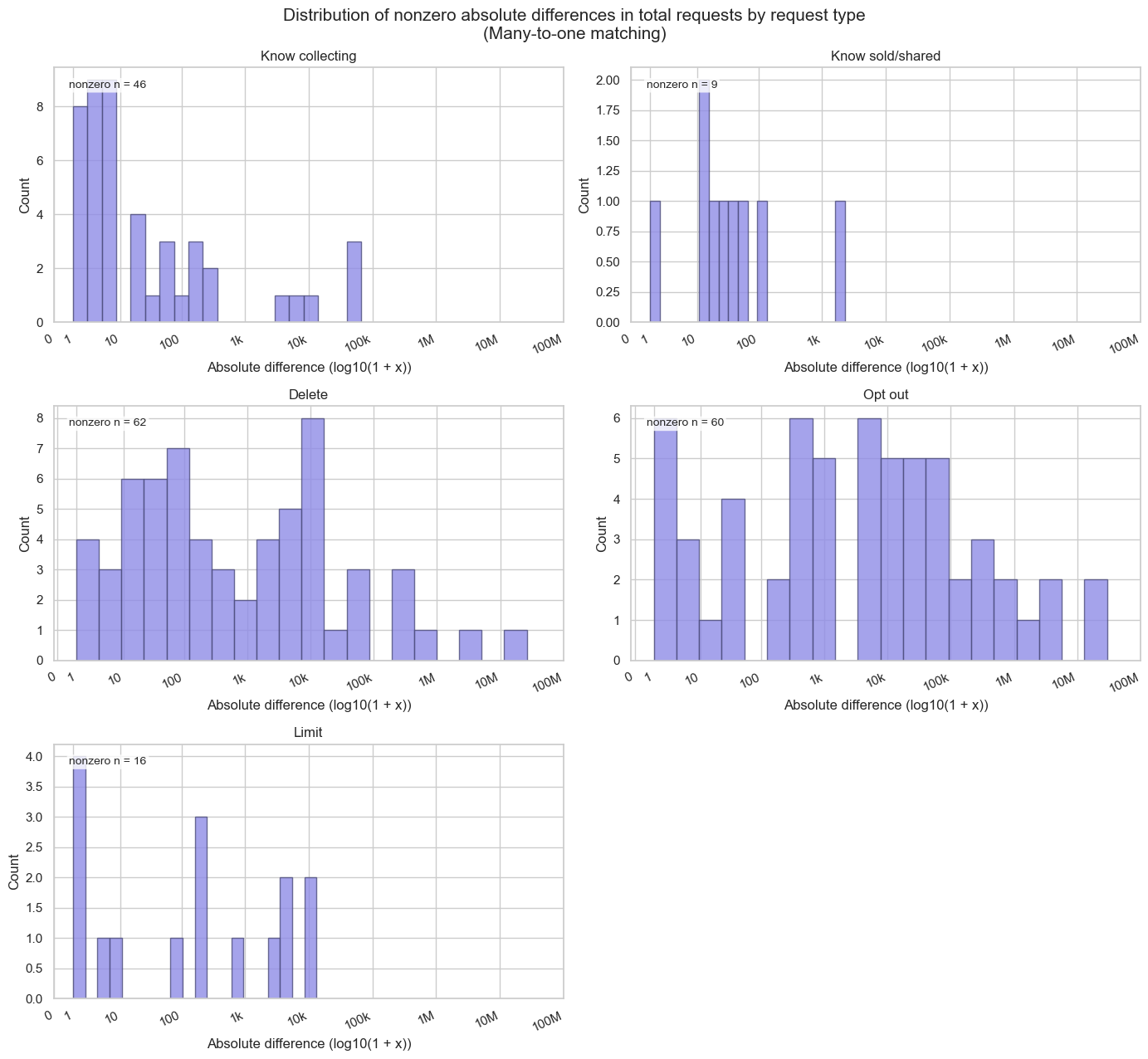}
    \caption{Distribution of nonzero absolute differences in total request counts by request type, using many-to-one broker matching. The figure includes only total request counts and excludes discrepancies in complied with and denied request counts.}
    \label{fig:absolute-diff-request-types}
    \Description{These figures represent the difference between our hand collected request metrics reported in 2025 to the metrics reported in the 2026 data broker registry, per request type}
\end{figure}

Figure~\ref{fig:absolute-diff-request-types} visualizes the distribution of nonzero absolute differences for total request counts by request type. This figure is limited to total request counts and excludes discrepancies in complied with and denied request counts; accordingly, its total count differs from the 494 numeric mismatch cells reported in Table~\ref{tab:registry-direction}. The figure shows that many total-count discrepancies are relatively small, but that deletion and opt-out requests include several large differences.

We also conducted a sensitivity analysis treating hand-coded ``none'' values as zero rather than as missing. Under this assumption, 365 matched brokers had at least one numeric discrepancy, and the direction of differences shifted strongly toward registry overreporting: 1,939 discrepancy cells (89.7\%) had higher values in the 2026 registry than in the hand-coded data, compared with 223 cells (10.3\%) where the registry value was lower. These results underscore the need for clearer reporting rules, machine-readable formats, and definitions that distinguish true zeros from missing or non-reported values.

Finally,  Table \ref{tab:2026-bivariate-analysis} shows how features of data brokers differ depending on if they received low or high request amounts (split on the median). We note that our analysis of 2025 registered brokers (Table 4 in the main body) found no significance in difference of number of requests based on features of type of data collected and which laws they must comply with. For 2026, there is a significant difference in collecting biometric data (data brokers reporting more requests tend to collect biometric data at higher rates) and other additional types of data.

\begin{table}[t]
\centering
\caption{Comparison of hand-collected request metrics to the 2026 California Data Broker Registry. Percentages are calculated over 457 matched brokers unless otherwise noted.}
\label{tab:registry-comparison}
\begin{tabular}{lrr}
\toprule
Comparison metric & Count & Percent \\
\midrule
Hand-coded metrics missing, but present in  2026 registry& 199 & 43.5\% \\
Both sources report at least one metric & 258 & 56.5\% \\
2026 registry reports more metric fields (including zero) & 412 & 90.2\% \\
2026 registry reports more nonzero metric fields& 320 & 70.0\% \\
Same metric coverage across both sources& 45 & 9.8\% \\
Same coverage and exact numeric agreement & 27 & 5.9\% \\
Brokers with at least one numeric mismatch & 115 & 25.2\% \\
Missing-only differences (metric reported to 2026 registry, but missing from hand-coded)& 315 & 68.9\% \\
\bottomrule
\end{tabular}
\end{table}

\begin{table}[t]
\centering
\caption{Direction and size of numeric mismatch cells. Percentages are calculated over 494 numeric mismatch cells where both sources reported a value. Difference size is reported as the absolute value of the hand-coded value minus the 2026 registry value.}
\label{tab:registry-direction}
\begin{tabular}{lrrrr}
\toprule
Difference type & Count & Percent & Median size & Mean size \\
\midrule
Registry value higher than hand-coded value & 271 & 54.9\% & 62 & 280{,}430 \\
Registry value lower than hand-coded value & 223 & 45.1\% & 37 & 419{,}326 \\
\bottomrule
\end{tabular}
\end{table}

\begin{table}[h]
\caption{Comparison of correlates between data brokers receiving low and high total numbers of requests for requests reported in 2026 data broker registry. We note that many of these questions were not asked in the 2025 registry, and thus, we do not have equivalent data for our original analysis. However, this table provides insight into the metrics found in the 2026 registry and their association with data broker collection characteristics.}
\label{tab:2026-bivariate-analysis}
\resizebox{\columnwidth}{!}{%
\begin{tabular}{@{}llllll@{}}
\toprule
 &
  \begin{tabular}[c]{@{}l@{}}Low requests received \\ (N=283)\end{tabular} &
   &
  \begin{tabular}[c]{@{}l@{}}High requests received\\ (N=283)\end{tabular} &
   &
   \\ \midrule
                                                                                    & Percentage   & SE                      & Percentage   & SE           & p-value       \\ \cmidrule(l){2-6} 
Collects data from minors                                                           & 4.2          & \multicolumn{1}{c}{1.2} & 2.1          & 0.9          & 0.23          \\
\begin{tabular}[c]{@{}l@{}}Collects security codes\\ for third parties\end{tabular} & 2.5          & 0.9                     & 2.1          & 0.9          & 1             \\
Collects government identification data                                             & 8.5          & 1.7                     & 6            & 1.4          & 0.33          \\
Collects citizen and immigration data                                               & 2.1          & 0.9                     & 1.8          & 0.8          & 1             \\
Collects data on union status                                                       & 0.7          & 0.5                     & 1.4          & 0.7          & 0.68          \\
Collects data on sexual orientation                                                 & 2.5          & 0.9                     & 3.9          & 1.1          & 0.47          \\
\textbf{Collects biometric data}                                                    & \textbf{3.5} & \textbf{1.1}            & \textbf{0.4} & \textbf{0.4} & \textbf{0.01} \\
Collects precise geolocation data                                                   & 20.1         & 2.4                     & 18.7         & 2.3          & 0.75          \\
Collects reproductive health data                                                   & 1.4          & 0.7                     & 1.1          & 0.6          & 1             \\
\textbf{\begin{tabular}[c]{@{}l@{}}Collects other additional name\\ and address information\end{tabular}} &
  \textbf{95.8} &
  \textbf{1.2} &
  \textbf{91.2} &
  \textbf{1.7} &
  \textbf{0.04} \\
Subject to FCRA                                                                     & 3.2          & 1                       & 2.8          & 1            & 1             \\
Subject to GLBA                                                                     & 4.6          & 1.2                     & 2.1          & 0.9          & 0.16          \\
Subject to IIPPA                                                                    & 0.4          & 0.4                     & 0            & 0            & 1             \\
Subject to CMIA                                                                     & 1.1          & 0.6                     & 1.1          & 0.6          & 1             \\
Subject to HIPAA                                                                    & 5.7          & 1.4                     & 4.6          & 1.2          & 0.70          \\
Sold consumer data to a foreign actor                                               & 3.9          & 1.1                     & 7.8          & 1.6          & 0.07          \\
Sold consumer data to federal government                                            & 10.2         & 1.8                     & 8.1          & 1.6          & 0.47          \\
Sold consumer data to state government                                              & 11           & 1.9                     & 6.7          & 1.5          & 0.10          \\
Sold consumer data to law enforcement                                               & 6.4          & 1.5                     & 3.2          & 1            & 0.11          \\
\begin{tabular}[c]{@{}l@{}}Sold consumer data to generative AI \\ company\end{tabular} &
  6.4 &
  1.5 &
  4.6 &
  1.2 &
  0.46 \\
From California                                                                     & 23.3         & 2.5                     & 18.4         & 2.3          & 0.18          \\ \bottomrule
\end{tabular}%
}
\end{table}

\clearpage
\subsection{Ranking Data Broker Opacity}
Using the design features we identified while reviewing data broker consumer request processes and their compliance with transparency requirements, we created two indexing frameworks to rank data brokers based on the most problematic factors we observed. The first uses compliance with transparency requirements, collection of sensitive data types, and friction features. Criteria are listed in Table 18. This scale is only applicable to the stratified sample of 250 brokers we analyzed for friction features in the consumer request process. The second ranking measures compliance with transparency requirements and collection of sensitive data types only and is applicable to all 522 data brokers. Table 19 presents this criteria. Both indexes are calculated out of a total value of 100, with higher values representing increased negative privacy and compliance impacts. Table 20 and 21 present the top 50 data brokers for each ranking method. These brokers represent the highest opacity scores, and thus are the least compliant to CCPA according to our measures. 

\vspace{1em}
\noindent
\begin{minipage}[t]{0.48\textwidth}
\centering
\small
\captionof{table}{Friction ranking scoring rubric.}
\label{tab:friction-rubric}
\begin{tabular}{l r}
\toprule
\textbf{Criterion} & \textbf{Points} \\
\midrule
\multicolumn{2}{l}{\textit{Friction}} \\
\quad No privacy policy & 8 \\
\quad Missing right to limit & 5 \\
\quad Missing right to know (selling/collecting) & 5 \\
\quad Missing right to opt out (do not sell) & 10 \\
\quad Missing right to delete & 10 \\
\quad Missing right to correct & 5 \\
\quad Broken link/email & 5 \\
\quad Excessive information & 5 \\
\quad Difficult to access / sensitive information required & 5 \\
\quad Identity verification required when unnecessary & 5 \\
\quad Separate/multiple forms & 2.5 \\
\quad CAPTCHA test & 2.5 \\
\midrule
\multicolumn{2}{l}{\textit{Compliance with Transparency Requirements}} \\
\quad Did not report any metrics on privacy policy & 14 \\
\midrule
\multicolumn{2}{l}{\textit{Sensitive Data Type}} \\
\quad Collects geolocation data & 6 \\
\quad Collects reproductive data & 6 \\
\quad Collects minors data & 6 \\
\midrule
\textbf{Total Possible Opacity Score} & \textbf{100} \\
\bottomrule
\end{tabular}
\end{minipage}%
\hspace{0.07\textwidth}
\begin{minipage}[t]{0.48\textwidth}
\centering
\small
\captionof{table}{Transparency requirements ranking scoring rubric.}
\label{tab:transparency-rubric}
\begin{tabular}{l r}
\toprule
\textbf{Criterion} & \textbf{Points} \\
\midrule
\multicolumn{2}{l}{\textit{Compliance with Transparency Requirements}} \\
\quad Does not post requests to delete & 20 \\
\quad Does not post requests to opt out & 20 \\
\quad Does not post requests to limit & 10 \\
\quad Does not post requests to know (collecting) & 10 \\
\quad Does not post requests to know (selling) & 10 \\
\midrule
\multicolumn{2}{l}{\textit{Sensitive Data Type}} \\
\quad Collects geolocation data & 10 \\
\quad Collects reproductive data & 10 \\
\quad Collects minors data & 10 \\
\midrule
\textbf{Total} & \textbf{100} \\
\bottomrule
\end{tabular}
\end{minipage}

\noindent
\begin{minipage}[t]{0.48\textwidth}
\centering
\footnotesize
\begin{longtable}{r l r}
\caption{Top 50 Data Brokers by Friction Opacity Score} \label{tab:friction} \\
\toprule
\textbf{\#} & \textbf{Data Broker} & \textbf{Opacity Score} \\
\midrule
\endfirsthead
\toprule
\textbf{\#} & \textbf{Data Broker} & \textbf{Opacity Score} \\
\midrule
\endhead
\midrule
\multicolumn{3}{r}{\textit{Continued\ldots}} \\
\endfoot
\bottomrule
\endlastfoot
    1 & Cengage Learning, Inc. & 61.0 \\
    2 & Trestle Solutions, Inc. & 59.0 \\
    3 & LexisNexis Risk Solutions FL Inc. & 49.5 \\
    4 & Dataskip & 49.0 \\
    5 & Lead411 Corporation & 46.5 \\
    6 & Market Force Corporation & 46.5 \\
    7 & Qualfon & 46.5 \\
    8 & SpyCloud, Inc. & 46.0 \\
    9 & Snovio Inc & 45.0 \\
    10 & AudiencePoint Inc. & 44.0 \\
    11 & VRTCAL Markets Inc & 43.5 \\
    12 & Comscore, Inc. & 43.5 \\
    13 & Uplead LLC & 42.5 \\
    14 & DealerSocket, LLC & 42.0 \\
    15 & Knower Tech USA, LLC & 41.0 \\
    16 & GRIN Technologies Inc. & 40.0 \\
    17 & 360 Media Direct & 39.0 \\
    18 & Accurate Append Inc. & 39.0 \\
    19 & Fushia Media, LLC. & 36.5 \\
    20 & LocateSmarter LLC & 36.5 \\
    21 & People Data Labs, Inc. & 36.5 \\
    22 & ModFx Labs Pvt Ltd & 36.0 \\
    23 & Sovrn, Inc. & 35.0 \\
    24 & LIZDEV INC. & 35.0 \\
    25 & Visual Visitor L.L.C. & 35.0 \\
    26 & USPEOPLESEARCH.COM, LLC & 35.0 \\
    27 & Traackr, Inc. & 34.0 \\
    28 & Trans Union LLC & 33.5 \\
    29 & Grassroots Analytics & 30.0 \\
    30 & Sharethrough Inc. & 30.0 \\
    31 & Windfall Data, Inc. & 30.0 \\
    32 & Growing Libraries, LLC & 30.0 \\
    33 & MedPro Systems & 30.0 \\
    34 & Family Tree Now, LLC & 30.0 \\
    35 & We Inform LLC & 30.0 \\
    36 & Next Wave Marketing Strategies, Inc & 30.0 \\
    37 & Marriott International, Inc. & 29.5 \\
    38 & TargetSmart Communications LLC & 29.0 \\
    39 & Revelio Labs, Inc. & 29.0 \\
    40 & Outward Media, Inc. & 29.0 \\
    41 & Qurium Solutions, Inc. & 29.0 \\
    42 & PaeDae, Inc. & 28.5 \\
    43 & Trans Union  Content Solutions LLC & 27.5 \\
    44 & Tru Optik Data Corp. & 27.5 \\
    45 & LightBox Parent, L.P. & 27.5 \\
    46 & MH Sub I, LLC & 27.5 \\
    47 & BH MARKETING GROUP LLC & 27.5 \\
    48 & Dun \& Bradstreet, Inc. & 27.0 \\
    49 & NetWise Data, LLC & 27.0 \\
    50 & SheerID, Inc. & 26.5 \\
\end{longtable}
\end{minipage}%
\hfill
\begin{minipage}[t]{0.48\textwidth}
\centering
\footnotesize
\begin{longtable}{r l r}
\caption{Top 50 Data Brokers by Opacity Score} \label{tab:transparency} \\
\toprule
\textbf{\#} & \textbf{Data Broker} & \textbf{Opacity Score} \\
\midrule
\endfirsthead
\toprule
\textbf{\#} & \textbf{Data Broker} & \textbf{Opacity Score} \\
\midrule
\endhead
\midrule
\multicolumn{3}{r}{\textit{Continued\ldots}} \\
\endfoot
\bottomrule
\endlastfoot
    1 & Lionshare Marketing, Inc & 100.0 \\
    2 & AlikeAudience, Inc. & 90.0 \\
    3 & HealthWise Data & 90.0 \\
    4 & Cengage Learning, Inc. & 90.0 \\
    5 & Informa USA Inc. & 90.0 \\
    6 & Datafy LLC & 80.0 \\
    7 & Quadrant Global Pte. Ltd. & 80.0 \\
    8 & 5X5 US, LLC & 80.0 \\
    9 & Reklaim Ltd. & 80.0 \\
    10 & Buyerlink Inc. & 80.0 \\
    11 & Blis Global Ltd & 80.0 \\
    12 & Venpath, Inc. & 80.0 \\
    13 & AutoWeb, Inc. & 80.0 \\
    14 & Direct Marketing Solutions, Inc. & 80.0 \\
    15 & Collective Data Solutions & 80.0 \\
    16 & Warmly, Inc & 80.0 \\
    17 & Mobile Technology Corporation & 80.0 \\
    18 & BH MARKETING GROUP LLC & 80.0 \\
    19 & CrawlBee Corp & 80.0 \\
    20 & True Blue Analytics LLC & 80.0 \\
    21 & Quad/Graphics, Inc. & 80.0 \\
    22 & HubSpot, Inc. & 80.0 \\
    23 & BDO GCI, LLC & 80.0 \\
    24 & Venntel, Inc. & 80.0 \\
    25 & Start.io Inc. & 80.0 \\
    26 & Unacast, Inc. & 80.0 \\
    27 & Hivestack Inc. & 80.0 \\
    28 & Valassis Communications, Inc. & 80.0 \\
    29 & Famous Birthdays LLC & 80.0 \\
    30 & Buildertrend Solutions, Inc. & 80.0 \\
    31 & iSpot.tv, Inc. & 80.0 \\
    32 & Veraset & 80.0 \\
    33 & Converge Direct, LLC & 80.0 \\
    34 & LightBox Parent, L.P. & 80.0 \\
    35 & Place Exchange, Inc. & 80.0 \\
    36 & IDG Communications Inc. & 80.0 \\
    37 & Irys, Inc & 80.0 \\
    38 & Grassroots Analytics & 80.0 \\
    39 & OnPoint Data Strategy & 80.0 \\
    40 & StackAdapt Inc. & 80.0 \\
    41 & Beeswax & 80.0 \\
    42 & CITYDATA Inc. & 80.0 \\
    43 & Acronymix LLC & 80.0 \\
    44 & Disco Technology Inc. & 70.0 \\
    45 & Semasio GmbH & 70.0 \\
    46 & RevOptimal, LLC & 70.0 \\
    47 & Traackr, Inc. & 70.0 \\
    48 & Findem, Inc. & 70.0 \\
    49 & UPS Capital Corporation & 70.0 \\
    50 & FIRST ORION CORP & 70.0 \\
\end{longtable}
\end{minipage}

Across the top 100 for both ranking methods, there are 20 data brokers that are in both top rankings for opacity scores. These brokers have high opacity scores, and thus are the least compliant with the CCPA given our measures. These are: 
\begin{itemize}
    \item 360 Media Direct
    \item Accurate Append Inc.
    \item BH MARKETING GROUP LLC
    \item Cengage Learning, Inc.
    \item Connected Investors, LLC
    \item Data Partners Inc.
    \item Datafy LLC
    \item Dataskip
    \item FIRST ORION CORP
    \item Grassroots Analytics
    \item Hivestack Inc.
    \item LightBox Parent, L.P.
    \item LocateSmarter LLC
    \item Reklaim Ltd.
    \item Semasio GmbH
    \item SheerID, Inc.
    \item Traackr, Inc.
    \item Trestle Solutions, Inc.
    \item VRTCAL Markets Inc
    \item iSpot.tv, Inc.
\end{itemize}

\end{document}